\documentclass[12pt]{article}

\usepackage{amsmath, amssymb, amsthm}
\usepackage{graphicx}
\usepackage{natbib} % for author-year citations
\usepackage{setspace} % for double spacing
\usepackage{hyperref} % for links in references

\usepackage[margin=1in]{geometry}

 \hypersetup{colorlinks,
 linkcolor=black,
 citecolor=blue,
 urlcolor=black
 }

 \usepackage{bbm}
 \usepackage{enumitem}

\usepackage{algorithm}
\usepackage{algorithmic}

\usepackage{xcolor}

\theoremstyle{plain}
\newtheorem{theorem}{Theorem}[section]
\newtheorem{proposition}[theorem]{Proposition}

\newtheorem{corollary}[theorem]{Corollary}
\theoremstyle{definition}

\theoremstyle{remark}
\newtheorem{remark}[theorem]{Remark}

\newcommand{\PP}{\mathbb{P}}
\newcommand{\EE}{\mathbb{E}}
\newcommand{\RR}{\mathbb{R}}
\newcommand{\calN}{\mathcal{N}}

\newcommand{\bX}{\boldsymbol{X}}

\newcommand{\bI}{\boldsymbol{I}}
\newcommand{\bfbeta}{\boldsymbol{\beta}}
\newcommand{\bY}{\boldsymbol{Y}}
\newcommand{\bepsilon}{\boldsymbol{\epsilon}}
\newcommand{\bgamma}{\boldsymbol{\gamma}}

\newcommand{\by}{\boldsymbol{y}}

\newcommand{\bZ}{\mathbf{Z}}

\newcommand{\bfeta}{\boldsymbol{\eta}}

\newcommand{\calD}{\mathcal{D}}
\newcommand{\bfb}{\boldsymbol{b}}

\DeclareMathOperator*{\argmin}{arg\,min}

\renewcommand{\vec}[1]{\boldsymbol{#1}}

\begin{document}

\title{Nonparametric Shrinkage Estimation in High Dimensional  Generalized Linear Models via Polya Trees}

\author{Asaf Weinstein \thanks{Department of Statistics, Hebrew University of Jerusalem, Email: asaf.weinstein@mail.huji.ac.il} 
\and Jonas Wallin \thanks{Department of Statistics, Lund University, Email: jonas.wallin@stat.lu.se}
\and
Daniel Yekutieli \thanks{Department of Statistics and OR, Tel Aviv University, Email: yekutiel@tauex.tau.ac.il}
\and Malgorzata Bogdan \thanks{Institute of Mathematics, University of Wroclaw, Email: malgorzata.bogdan@uwr.edu.pl}}

\date{}

\maketitle

\begin{abstract}
Regularization in fitting regression models has been a highly active topic of research in the past few decades, but most of the existing methods are designed for particular situations, e.g. for the case of a sparse coefficient vector. 
We consider the problem of designing {\em universally} optimal regularized estimators in a given generalized linear model with fixed effects. 
First, we propose as a contender the Bayes estimator against an {\em ideal} prior that assigns equal mass to every permutation of the fixed coefficient vector, thus depending on the true coefficients only through their empirical CDF. 
We prove some optimality properties of this oracle estimator in both the frequentist and Bayesian frameworks. 
To compete with the oracle estimator, we posit a hierarchical Bayes model where the individual coefficients are modeled as i.i.d.~draws from a common distribution $\pi$, which is in turn assigned a Polya tree prior to reflect indefiniteness. 
We demonstrate in examples that the posterior mean of $\pi$ under the postulated model adapts nonparametrically to the empirical CDF of the true coefficients. 
Correspondingly, the posterior means of the coefficients themselves are used to mimic the ideal estimator. 
Numerical experiments show that our method has better estimation and prediction accuracy compared to various parametric and nonparametric alternatives, from relatively standard $L_p$-regularized estimators to modern penalized-likelihood and Bayesian estimators for high dimensional regression.
\end{abstract}


\section{Introduction}\label{sec:intro}

Supervised learning problems nowadays often entail fitting complex models with thousands or even hundreds of thousands of parameters, sometimes exceeding the number of observed cases. 
Correspondingly, incorporating regularization in the training process is paramount to controlling overfitting and enabling generalization to new examples. 
% While there is agreement on the need for some kind of regularization, 
Still, {\em what} exact form of regularization is most adequate for a given problem is generally far from obvious. 
%{\em what} particular form of regularization one should use under a given model, is far less obvious. 
In this paper we consider the problem of designing an {\em optimal} regularizer in a setting where the observations $(\bX_i, Y_i),\ i=1,...,n$, follow a generalized linear model (GLM), 
\begin{equation}\label{eq:glm}
Y_i \overset{ind}{\sim} f(y_i;\eta_i, \psi),\ \ \ \EE_{\eta_i}[Y_i] = g^{-1}(\eta_i),\ \ \ \eta_i = \bX_i^\top \bfbeta,
\end{equation}
where $\bX_i\in \mathbb{R}^p$ are fixed and known covariate vectors, and $f$ is a given likelihood function {corresponding to an exponential family of distributions}. 
The main parameter vector $\bfbeta = (\beta_1,...,\beta_p)^\top$, and the (optional) nuisance parameter $\psi$ are unknown, and the task is, in general, to estimate $\bfbeta$ under a specified loss function. 
We will consider mainly the squared loss in estimating the coefficient vector $\bfbeta$ or the linear predictor $\bfeta = (\eta_1,...,\eta_n)^\top = \bX\bfbeta$, where $\bX = [\bX_1,...,\bX_n]^\top \in \mathbb{R}^{n\times p}$, but the methods to be described in the sequel apply in principle to any loss function. 

A standard approach to obtain a regularized estimator for $\bfbeta$ in \eqref{eq:glm}, is to maximize a {\it penalized} version of the likelihood, 
\begin{equation}\label{eq:pen-lik}
\widehat{\bfbeta} = \arg \max_{\bfbeta\in \RR^p} \sum_i \log f(Y_i; \eta_i, \psi) + \mathcal{P}_{\lambda}(\bfbeta),
\end{equation}
where $\mathcal{P}_{\lambda}(\bfbeta)$ is a regularization function indexed by $\lambda$ and specified in advance. 
We assume here that for any two values $\psi,\psi'$, the estimator $\widehat{\bfbeta}$ calculated under $\lambda, \psi$ is the same as that calculated under $\lambda', \psi'$ for some $\lambda'$, so that the objective is in a sense independent of $\psi$; this is the case, e.g., in the linear model ($g(\mu) = \mu,\ \psi=\sigma$), and, trivially in the logistic model ($g(\mu) = \ln (\mu/(1-\mu))$), which has no nuisance parameters. 
Penalized-likelihood estimators \eqref{eq:pen-lik} shrink by balancing the log-likelihood against `suitably disciplined' values of $\bfbeta$, to use the terminology of \cite{rovckova2018spike}. 
There is plenty of modern work proposing and analyzing penalized-likelihood estimators for GLMs, far beyond the relatively standard options of $L_p$ penalties. 
These include, for example, the convex methods of 
\citet[][OSCAR]{OSCAR} and \citet[][SLOPE]{bogdan2015slope}, which promote both variable selection and parameter tying,  or the nonconvex methods of \citet[][SCAD]{fan2001variable} and   \citet[][MCP]{zhang2010nearly}, that simultaneously perform selection and shrinkage estimation with carefully designed nonconcave penalties for attenuating  bias. 
% \color{cyan}\cite{salehi2019impact} studied a convex regularizer that promotes a desired (known) structure in a logistic regression model; 
% and \cite{xu2017generalized} studied regularization based on distance penalties deriving from prespecified constraints. 
% \cite{jain2014iterative} revisit the Iterative Hard Thresholding algorithm of \cite{blumensath2009iterative} and provide analysis in a pragmatic high dimensional setting. 
% {\it Gosia: I think we may remove these references to ML papers. I do not think they are so well cited.}
% \color{black}

From a Bayesian perspective, the penalized-likelihood estimator \eqref{eq:pen-lik} can be viewed as a maximum {\em a posteriori} (MAP) estimator under the (possibly improper) prior $\pi(\bfbeta|\lambda) = \exp(\mathcal{P}_{\lambda}(\bfbeta))$. 
%In that sense, specifying an appropriate penalty function is equivalent to specifying an appropriate prior. 
The Bayesian viewpoint is often more convenient because {\em a priori} knowledge about $\bfbeta$ can be incorporated more directly into the model. 
This connection to Bayes estimators has been exploited in many existing papers that propose different penalty functions by a careful choice of a prior. 
%, mostly for the linear model. 
The majority of modern work concentrates on recovering a sparse coefficient vector, typically by employing various parametric {\em spike-and-slab} models. 
%building sparsity into the prior as a ``spike" about zero, while a ``slab" component generally has some restricted parametric form. 
%, and ultimately use {\it parametric} families of priors for regularization. 
%\cite{johnstone2004needles} advocate options with heavier tails than Normal, including the Laplace prior, while concentrating on the simpler, normal means model. 
\cite{george2000calibration} consider parametric classes of priors suitable for model selection and sparsity, and propose to estimate the hyperparameters from the data, resulting in empirical Bayes (EB) estimates. 
The alternative EB methods of \cite{yuan2005efficient} offer better computational efficiency. 
\cite{andersen2017bayesian} generalize the spike-and-slab prior to situations where the coefficient vector has a spatio-temporal structure. 
The Spike-and-slab Lasso method of \cite{rovckova2018spike} employs a (parametric) modification of the $L_1$ penalty, that similarly promotes sparsity but is able to reduce the bias of the Lasso. 
\cite{jiang2019adaptive} extend the methods from \cite{rovckova2018spike} , proposing a Bayesian counterpart of SLOPE.
%\citep[SLOPE;][]{bogdan1969statistical, jiang2019adaptive}. 
The estimator of \cite{carvalho2010horseshoe} uses the horseshoe prior, offering robustness to unknown sparsity level and the handling of large signal components. 
Many other Bayesian estimators have been proposed over the years, the vast majority of which again ultimately using {\it parametric} families of priors for regularization. 

We adopt the Bayesian viewpoint discussed above, postulating that $\beta_j$ are i.i.d.~draws from an unknown prior $\Pi$, but we make no assumptions at all on this prior, in particular there is no assumption that it is `sparse'. 
We propose a flexible hierarchical modeling scheme to estimate  this prior non-parametrically, specifically, to adapt to the ``true" prior (we discuss in Section \ref{sec:oracle} what this means when $\beta_j$ are  fixed) we model $\Pi$ itself as a random realization from some distribution $P$. 
In additiong to its nonparamtric nature, we want $P$ to be `noninformative' so it can be learned completely from the data. 
There is more than one option to choose such a $P$, for example taking it to be the distribution of a Dirichlet process \citep{ferguson1973bayesian} is a common choice in existing nonparametric Bayes literature. 
Here we propose to use for $P$ a {\em Polya tree} distribution. 
Polya trees belong to a class of tail-free distributions introduced by Lavine \citep{lavine1992some, lavine1994more} as a generalization of Ferguson's Dirichlet process  which, in particular, allows the random probability measure to be supported on continuous distributions, and maintains tractability. 
As detailed in Section \ref{sec:method}, Polya trees admit conjugacy (closure) properties, as does the Dirichlet process, which make it a convenient choice in practice. 

In essence, our approach is is a fully Bayes alternative to the nonparametric empirical Bayes approach, which regards $\Pi$ as a {\em fixed} and unknown member of a rich family of distributions specified in advance. 
% the empirical Bayes modeling in \cite{kim2022flexible}, which regards $\Pi$ as a {\em fixed} and unknown member of a rich (but still parametric) family of distributions specified in advance. 
The hierarchical Bayes approach in random-effects models has been mentioned by \citet{robbins1963empirical} in a nonparametric context, and is perhaps more familiar in parametric contexts \citep{efron2012large}. 
Both the EB and the fully Bayes approaches are ultimately intended to allow $\Pi$ to be learned (``deconvolved") using {\em all} of the observations $(\bX_i, Y_i)$. 
In \cite{kim2022flexible} the EB approach is pursued, modeling $\Pi$ as a mixture of zero-mean normal distributions with different (prespecified) scale parameters. 
In estimating these hyperparameters, \cite{kim2022flexible} use a variational approach to handle intractability of the posterior of $\bfbeta$. 
More precisely, the posterior of $\bfbeta$ is approximated by a product distribution, optimized within a prespecified family $\mathcal{Q}$ to best fit the true posterior in the sense of minimizing the Kullback-Leibler divergence; this makes it possible to leverage results from the simpler and well-studied sequence model. 
Avoiding such mean-field approximations of the posterior distribution is what we view as one of the advantages of the MCMC approach proposed in the current article.
% this is a fully Bayes alternative to the nonparametric EB approach in  \cite{kim2022flexible}, which treats $P$ as a fixed (deterministic), unknown member in a rich parametric family of distributions specified in advance. 

Polya trees have some history in application to nonparametric Bayes problems, however existing work generally restricts attention {to the `separable' (or `sequence') case, where the likelihood of each observation $Y_i$ depends on a separate parameter $\theta_i$ and there is no assumption of any relationship between the $\theta_i$'s.} 
%the parameters $\eta_i$ in the likelihood function $f(y_i;\eta_i, \psi)$ are ``free", i.e., there is no known relationship between them. 
This includes applications to EB problems in a `classic' setting \citep{antoniak1974mixtures, berry1979empirical, lavine1994more} and to nonparametric regression \citep{antoniak1974mixtures} and density estimation \citep[][and references therein]{castillo2017polya}. 
The novelty of the proposed methodology is in using a Polya tree prior with the model \eqref{eq:glm}, in which the $\eta_i$'s are all related through $\bfbeta$, so that the likelihood of each $Y_i$ depends on the same set of parameters $\beta_1,...,\beta_p$. 
This `non-separable' case introduces substantial complications that are not readily addressed by existing methods.

We have focused above on an overview of the proposed methodology, but an important aspect of this work---and something we find to be largely missing in related literature---is providing theoretical justification for the  nonparametric Bayes approach even when working in a strictly frequentist setting. 
Thus, after describing our hierarchical  Bayes model more precisely, in Section \ref{sec:oracle} we motivate our methods as pursuing an {\em oracle} estimator, defined to be the optimal estimator which knows the true vector $\bfbeta$ but is restricted in to use this knowledge {\em symmetrically}, i.e., without giving {\em a priori} preference to any of the orderings of $\bfbeta$. 
This basically means that only information about the unordered vector $\bfbeta$---equivalently, the empirical distribution of the true coefficients $\beta_1,...,\beta_p$---is available to the oracle. 
Importantly, the trivial solution $\hat{\bfbeta} = \bfbeta$ violates this condition, and therefore is eliminated. 
%\citep{lindley1972bayes}, 
%intuitively can be thought of as having access to the true $\bfbeta$ up to ordering. 
We provide supporting theoretical analysis showing that our oracle estimator has certain optimality properties in both a frequentist and a Bayesian framework, 
{and explain how this analysis is relevant to the Polya-tree based methodology proposed in this paper.}

The rest of the article is organized as follows. 
Section \ref{sec:method} presents our methodology by describing a hierarchical  model that includes a Polya tree prior on univariate distributions. 
After presenting an analysis of the oracle estimator in Section \ref{sec:oracle}, we provide in Section \ref{sec:simulations} results from a simulation in a logistic regression model. 
In Section \ref{sec:real-data} we apply our method to analyze real data of polygenic inheritance. 
We cocnlude in Section \ref{sec:discussion} with some remarks and directions for further research. 

\section{Methodology}
\label{sec:method}

%Working under the model \eqref{eq:glm}, we 
{Our approach starts by positing a hierarchical model on the observed data. 
Thus, we suppose 
\begin{equation}\label{eq:prior-iid}
\beta_1,...,\beta_p \overset{iid}{\sim} \Pi, %,\ \ \ \ \ \ \ \ \psi\sim h, 
\end{equation}
where the distribution $\Pi$ itself is modeled as random, 
\begin{equation}\label{eq:prior-pi}
\Pi \sim P
\end{equation}
for a specified `noninformative' distribution $P$. 
We emphasize that, unless otherwise indicated, \eqref{eq:glm} is the only modeling assumption for the observed data, so \eqref{eq:prior-iid}, \eqref{eq:prior-pi}, or any of the other suppositions that follow, are merely used to facilitate a shrinkage estimator. 
%the parameters $\beta_j$ are not actually random. 
To complete the Bayesian model, we further assume 
\begin{equation}\label{eq:prior-psi}
\psi\sim h, 
\end{equation}
independently of $\bfbeta$, where $h$ is a fully specified vague prior on the nuisance parameter $\psi$ (as usual, $h$ is allowed to depend on the likelihood $f$ in \eqref{eq:glm}). 
If we fix a loss function and regard $h$ as given, the choice of $P$ determines the Bayes rule for $\bfbeta$, which will be used as a regularized estimator; e.g., for squared loss, this is the posterior mean of $\bfbeta$ under $P$ and $h$.}

Parametric choices of $P$, e.g.~Gaussian, yield Bayes rules that generally resemble parametric EB estimators, some examples of which were mentioned in the Introduction. 
%Here, instead, we want to allow the model to learn a {\em completely unspecified} prior $P$, in particular, we want to adapt to sparse or dense $\bfbeta$ depending on the actual underlying signal. 
Instead, to allow our hierarchical model to learn a completely unknown $\Pi$, we take $P$ to be a {\em Polya tree} distribution. 
Polya trees, introduced by Ferguson, belong to a class of tail-free distributions on random probability measures that generalize Dirichlet Processes while maintaining tractability \citep{ferguson1973bayesian, ferguson1974prior}. 
Under the hierarchical Bayes model given by \eqref{eq:glm}, \eqref{eq:prior-iid}, \eqref{eq:prior-pi} and \eqref{eq:prior-psi}, we propose to use posterior sampling to provide inference for $\bfbeta$. 
This is carried out with a Gibbs sampling algorithm, which we construct to take advantage of conjugacy properties of Polya trees when conditioning on certain parts of the unobserved variables. 
% \revv{An implementation of the Gibbs sampler is available as a preliminary \texttt{R} package at \url{https://github.com/JonasWallin/hBayes}. }

We proceed with a more precise  description of the hierarchical model above, which we will also refer to as the {\em generative} model, to distinguish it from the frequentist model \eqref{eq:glm}, 
and then explain how we use it to provide (approximate) inference for $\bfbeta$. 
%To clarify, we use the term ``generative" to distinguish the Bayes model resulting form \eqref{eq:prior-iid}-\eqref{eq:prior-pi}, which we only hypothesize in order to produce a shrinkage estimator, from the {\em true} model \eqref{eq:glm} where $\bfbeta$ is regarded as fixed. 
% The main component in the generative model is the specification of $\Pi$ as a Polya tree distribution, which we describe below. 

\vspace{10pt}

\noindent {\bf The finite Polya tree model}. 
The {$L$-level finite Polya tree} (FPT) model generates distributions $\Pi$ with piecewise constant density functions on a dyadic partition of ${\cal I}_0 = ( a_{min}, a_{\max}]$, 
corresponding to a fixed endpoints vector $\vec{a} := (a_{min} = a_0 \le a_1 \le  \cdots \le a_{2^L-1}  \le  a_{2^L} = a_{max})$. 
The dyadic partition consists of subintervals
${\cal I}_{l,i} = (a_{(i-1) \cdot 2^{L-l}}, a_{i \cdot 2^{L-l}}]$,  for $l = 1 \cdots L$ and $i = 1 \cdots 2^l$.
The parameters of the FPT model are the Beta parameters $( \alpha_{l, i}, \beta_{l, i})$, 
corresponding to subintervals ${\cal I}_{l-1,i}$ for $l = 1 \cdots L$ and $i = 1 \cdots 2^{l-1}$.
 The  FPT model has the following components.
 \begin{itemize}
\item[ {I. }] {\em Independent Beta random variables.}
A vector $\boldsymbol{\phi} = ( \phi_{1,1} \cdots \phi_{L, 2^{L-1}})$ of independent Beta random variables $\phi_{l, i} \sim Beta ( \alpha_{l, i}, \beta_{l, i})$, specifying conditional subinterval probabilities for the dyadic partition. 
Specifically, $\PP({\cal I}_{1, 1} | {\cal I}_0  ) = \phi_{1, 1}$ and $\PP({\cal I}_{1,2}  | {\cal I}_0 )  = 1 - \phi_{1, 1}$, and, 
for $l = 2 \cdots L$ and $i = 1 \cdots 2^{l-1}$, $\PP({\cal I}_{l, 2 \cdot i - 1} | {\cal I}_{l-1, i}) = \phi_{l, i}$, $\PP({\cal I}_{l, 2 \cdot i } | {\cal I}_{l-1, i}) = 1 - \phi_{l, i}$.

\item[ {II. }] {\em Subinterval probabilities.}  The subinterval probabilities vector $\vec{\pi}$ has elements $\PP({\cal I}_{l,i})  = \pi_{l, i}$, which 
are products of the conditional subinterval probabilities: 
$\pi_{1, 1} =  \phi_{1, 1}$, $\pi_{1, 2} = 1 - \phi_{1, 1}$, and, 
for $l = 2,...,L$ and $i = 1,...,2^{l-1}$, 
$\pi_{l,2 \cdot  i -1} = \phi_{l,  i} \cdot \pi_{l-1,i}$ and $\pi_{l,2 \cdot  i} = (1- \phi_{l,  i}) \cdot \pi_{l-1,i}$.
 
\item[ {III. }] {\em Step function PDF.}
The step function density $\Pi$ is specified conditionally on the vector $\vec{\pi}_L = (\pi_{L, 1},..., \pi_{L, 2^L})$ of subinterval probabilities at level $L$,  
\begin{equation}\label{def-step}
    \begin{aligned}
\Pi (  \beta |   \vec{\pi}_L )  =  
\pi_{L,1} \cdot  \mathbbm{1}_{L,1} (\beta) /  (a_1 - a_0) +  \cdots 
+
\pi_{L, 2^L} \cdot \mathbbm{1}_{L, 2^L} (\beta)  / (a_{2^L}  - a_{2^L-1}),
\end{aligned}
\end{equation}
for $\beta \in {\cal I}_0$, where $\mathbbm{1}_{L, i} (\beta)$ is the indicator function corresponding to ${\cal I}_{L,i}$. 
 \end{itemize}
%With the definitions above, we provide posterior inference assuming that the observed data $\bY$ has been generated from the following generative model: 
%is a realization from the generative model: 
Taking $P$ to be a level-$L$ FPT, this results in the following generative model: 
\begin{enumerate}[label=\roman*)]
\item draw ${\psi} \sim h ({\psi} )$. 
\item draw $\Pi$ from the FPT model with $\phi_{l,i} \sim Beta(1,1)$.  
\item draw $\beta_j \sim \Pi$, i.i.d., for $j = 1,...,p$. 
\item 
% given $\bfbeta = (\beta_1,...,\beta_p)^\top$ and $\psi$, 
draw $\vec{Y}$ from \eqref{eq:glm}. 
% \vspace{-0.1cm}
% \begin{enumerate}
% \item Draw ${\psi} \sim g ({\psi} )$. \vspace{-0.05cm}
% \item Draw $\Pi$ from the FPT model with $\phi_{l,i} \sim Beta(1,1)$.
% \vspace{-0.05cm}
% %\item Generate $\Pi (  \cdot |   \vec{\pi}_L ) $ from the FPT model with $\phi_{l,i} \sim Beta(1,1)$.
% \item Draw $\beta_j \sim \Pi$, i.i.d., for $j = 1,...,p$. 
% \vspace{-0.05cm}
% \item Given $\bfbeta = (\beta_1,...,\beta_p)^\top$ and $\psi$, draw $\vec{Y}$ from \eqref{eq:glm}. 
% %, where $f(\by;\bfeta, \psi)$ is the likelihood in \eqref{eq:glm}. 
%  \end{enumerate}
%  % \end{definition}
\end{enumerate}

The posterior distribution of $\boldsymbol{\Theta}=\{\psi, \boldsymbol{\beta},\boldsymbol{\phi}\}$
is sampled by Gibbs sampling as described in Algorithm \ref{alg:seq}. 
The single-site Metropolis Hastings (MH) algorithm detailed in section 
% B
\ref{sec:sample_beta}
of the Appendix, is used to sample each component $\beta_j$ of $\boldsymbol{\beta}$ conditionally on the remaining components of $\boldsymbol{\beta}$, 
and on $\psi$, $\boldsymbol{\phi}$ and $\by$. 
To sample $\psi$ conditionally on $(\bfbeta, \by)$, one can use a MH step, unless the prior is (conditionally) conjugate, in which case direct sampling is easy. 
For instance, for a Gaussian linear model, $\psi=\sigma^2$ and if an inverse-Gamma prior is used then the conditional posterior is also a inverse-Gamma. 
% (DY: I THINK THAT IN EXPRESSION \eqref{eq:glm} WE NEED TO CHANGE THE EXPRESSION FOR THE LIKELIHOOD  TO  $f(\boldsymbol{y}; \boldsymbol{\beta}, \psi)$).
Finally, for sampling $\vec{\phi} | \boldsymbol{\beta}, \by := \vec{\phi} |\, \boldsymbol{\beta}$, 
let
$$
N_{l,i} (\bfbeta) = \#\{ j : \beta_j \in {\cal I}_{l, i} \}, 
$$
and let  $\boldsymbol{N} = ( N_{1,1},..., N_{L,2^L})$. 
\cite{ferguson1974prior} has already noted the conjugacy of the FPT model, namely, that $\vec{\phi} |\, \boldsymbol{\beta}$  is a FPT with updated hyper-parameter values, 
% \vspace{-0.2cm}
\begin{equation} \label{no-noise-pst}
 \phi_{l, i} | \boldsymbol{N}  \sim Beta( 1 + N_{l, 2 \cdot i - 1}, 1 + N_{l, 2 \cdot i}).
 \end{equation}
% \vspace{-0.3cm}
% Algorithm \ref{alg:seq} describes the Gibbs sampling algorithm.

% \begin{algorithm}[htbp]  % Add optional float specifier
%     \SetAlgoLined
%     \SetKwInOut{Set}{Set}
%     \SetKwInOut{Input}{Input}
%     \SetKwInOut{Output}{Output}
    
%     \Set{number of iterations $G$, number of levels $L$, endpoints vector  $\vec{a}$,  initial values $\vec{\Theta}^{(0)}$ }
%     \Input{$\vec{y}$, $\vec{X}$}
%     \Output{posterior samples $\vec{\Theta}^{(1)}  \cdots, \vec{\Theta}^{(G)}$}
    
%     \For{$g = 1, \cdots, G$}{
%         \For{$j = 1, \cdots, p$}{
%             Sample $\beta_j^{(g)}|\psi^{(g-1)}, \vec{\phi}^{(g-1)}, \vec{y},$ 
%             $(\beta^{(g)}_1, \cdots, \beta^{(g)}_{j-1}, \beta^{(g-1)}_{j+1}, \cdots, \beta^{(g-1)}_{p})$ by MH
%         }
%         Sample $\psi^{(g)}|\vec{\beta}^{(g)}, \vec{y}$ \\
%         Sample $\vec{\phi}^{(g)}|\vec{\beta}^{(g)}$ using  \eqref{no-noise-pst}
%     }
%     \caption{Gibbs sampler for GLM}  % Ensure no extra spaces before \caption
%     \label{alg:seq}
% \end{algorithm}

\begin{algorithm}
\caption{Gibbs Sampler for GLM}
\label{alg:seq}
\begin{algorithmic}[1]
\STATE \textbf{Set:} number of iterations $G$, number of levels $L$, endpoints vector $\vec{a}$, initial values $\vec{\Theta}^{(0)}$
\STATE \textbf{Input:} $\vec{y}$, $\vec{X}$
\STATE \textbf{Output:} posterior samples $\vec{\Theta}^{(1)}, \ldots, \vec{\Theta}^{(G)}$
\FOR{$g = 1$ to $G$}
  \FOR{$j = 1$ to $p$}
    \STATE Sample $\beta_j^{(g)} \mid \psi^{(g-1)}, \vec{\phi}^{(g-1)}, \vec{y}, \beta_1^{(g)}, \ldots, \beta_{j-1}^{(g)}, \beta_{j+1}^{(g-1)}, \ldots, \beta_p^{(g-1)}$ by MH
  \ENDFOR
  \STATE Sample $\psi^{(g)} \mid \vec{\beta}^{(g)}, \vec{y}$
  \STATE Sample $\vec{\phi}^{(g)} \mid \vec{\beta}^{(g)}$ using Eq.~\eqref{no-noise-pst}
\ENDFOR
\end{algorithmic}
\end{algorithm}

\section{An oracle shrinkage rule}
\label{sec:oracle}

In this section we provide some theoretical support for the proposed approach. 
We will need to distinguish, only in this section, between the true, fixed value of the coefficient vector in \eqref{eq:glm}, which we denote here $\bfbeta^*$, and a general (potential) value, which we denote $\bfbeta$. 
% Hence, $\bfbeta^*$ is  a fixed unknown parameter. 
While our postulated Bayes model produces a legal estimator, because it is a function of the data only, it is not obvious what object we are actually estimating in terms of the original  model \eqref{eq:glm}, or why such an object may be a good target to pursue. 
To give an intuitive argument, the connection mentioned earlier between the fully Bayes approach and the EB approach---both facilitating estimation of $\Pi$ from the entire set of observations---suggests interpreting 
the Bayes rule for $\bfbeta$ under the hierarchical model postulated in the previous section as an estimator of an {\em oracle} Bayes rule which uses the prior
\begin{equation}\label{eq:oracle-prior-iid}
\beta_1,...,\beta_p \overset{i.i.d.}{\sim} \Pi^*_p,
\end{equation}
where $\Pi^*_p = \frac{1}{p}\sum_{j=1}^p \delta_{\beta^*_{j}}$ is the distribution that puts mass $1/p$ on each of $\beta^*_1,...,\beta^*_p$, i.e., the {\em empirical distribution} of the components $\beta^*_{1},...,\beta^*_{p}$ of $\bfbeta^*$. 
%Hence, $\Pi^*_p $ depends on .
When $p\to \infty$, the i.i.d.~prior \eqref{eq:oracle-prior-iid} is in turn expected to be similar, in a sense, to the {\em exchangeable} prior 
\begin{equation}\label{eq:prior-oracle}
    \bfbeta\sim \widetilde{\Pi}^*_p,
\end{equation}
where $\widetilde{\Pi}^*_p$ denotes the uniform distribution on all $p!$ permutations of the vector $\bfbeta^*$; see e.g.~\cite{diaconis1980finite} and \cite{hannan1955asymptotic, greenshtein2009asymptotic}. 
Note that both $\Pi^*_p$ and $\widetilde{\Pi}^*_p$ depend on $\bfbeta^*$ only up to ordering, in other words, only through the multiset\footnote{a multiset accounts for duplicates, unlike a set.} $\{\bfbeta^*\}:= \{\beta^*_1,...,\beta^*_p\}$. 
Note also that these two priors differ in that $\Pi^*_p$ samples $p$ draws from the empirical distribution of the $\beta_j$'s with replacement, while $\widetilde{\Pi}^*_p$ samples $p$ draws without replacement, explaining intuitively why we expect them to be `similar' for large $p$. 
Putting these pieces together, the Bayes rule under the prior \eqref{eq:prior-iid}-\eqref{eq:prior-pi} can be understood intuitively as pursuing the {\em oracle} Bayes rule under the prior \eqref{eq:prior-oracle}, let us denote it $\widehat{\bfbeta}_{ol}$ (a formal definition will follow). 
%Instead of studying the estimator resulting from our method itself, \revv{which } 

In this section we study properties of $\widehat{\bfbeta}_{ol}$, which, by the argument just presented, we view as the 
%population-level 
target of the estimator produced by the proposed hierarchical Bayes method; note that $\widehat{\bfbeta}_{ol}$ is perfectly well defined in terms of the original frequentist model \eqref{eq:glm}. 
We present two separate results: the first is limited to the Normal linear model, and says that, if $\bX$ satisfies some conditions, $\widehat{\bfbeta}_{ol}$ minimizes the (point) {\em risk} 
$
R(\bfbeta^*, \widehat{\bfbeta}) := \mathbb{E}_{\bfbeta^*} L(\bfbeta^*, \widehat{\bfbeta}), 
$
among all members in a very natural class of estimators, by appealing to permutation invariance considerations. 
The second result holds for any GLM and any matrix $\bX$, but its optimality guarantees are weaker. 
More specifically, it says that, under {\em any exchangeable} prior $\widetilde{\Pi}$ on $\bfbeta$, the estimator $\widehat{\bfbeta}_{ol}$ is optimal in terms of the {\em Bayes risk}, 
\begin{equation}\label{eq:bayes-risk}
    r(\widetilde{\Pi},\widehat{\bfbeta}) := \int R(\bfbeta, \widehat{\bfbeta})\widetilde{\Pi}(d\bfbeta). 
\end{equation}
To formalize things, first define the {\em oracle Bayes rule} to be the minimizer of \eqref{eq:bayes-risk} when $\widetilde{\Pi}=\widetilde{\Pi}^*_p$, 
\begin{equation}\label{eq:oracle-bayes}
    \widehat{\bfbeta}_{ol}(\bY) 
=
\argmin_{\bfb\in \RR^p} \EE_{\widetilde{\Pi}^*_p}[L(\bfbeta, \bfb)\lvert \bY], 
\end{equation}
the subscript on the expectation operator indicating that the (posterior) expectation of $\bfbeta$ is computed under the prior $\widetilde{\Pi}^*_p$. 
For simplicity, we will assume throughout this section that the nuisance parameter $\psi$ is known; the arguments that follow can be adapted to the case of unknown $\psi$, but we avoid this so as not to distract from the main ideas. 

Our first result is stated under the Gaussian linear model, 
$\bY\sim \calN_n(\bX\bfbeta^*, \psi^2\bI)$,  assuming $\bX\in \RR^{n\times p}$ has full column rank. 
In that case, a sufficient statistic is
\begin{equation}\label{eq:model-suff}
\bZ = (\bX^\top \bX)^{-1}\bX^\top \bY \sim \mathcal{N}_p(\bfbeta^*, \psi^2(\bX^\top \bX)^{-1}), 
\end{equation}
so that it makes sense to restrict attention to estimators of $\bfbeta^*$ that depend on $\bY$ only through $\bZ$. 
Now consider matrices $\bX$ s.t.,~for some $-1 < \rho < 1$, 
\begin{equation}\label{eq:gram}
\bX^\top \bX  \propto (1-\rho)\bf{I} + \rho \bf{1}\bf{1}^\top. 
%\bX^\top \bX \propto 
%\left[\begin{array}{ccccc}
%1 & \rho & \rho & \cdots & \rho \\
%\rho & 1 & \rho & \cdots & \rho \\
%\rho & \rho & 1 & \cdots & \rho \\
%\vdots & \vdots & \vdots & \ddots & \vdots \\
%\rho & \rho & \rho & \cdots & 1
%\end{array}\right]. 
\end{equation}
In this case, $\bZ-\bfbeta^*\sim \mathcal{N}_p(\boldsymbol{0}, \psi^2(\bX^\top \bX)^{-1})$ is an {\em exchangeable} random vector. 
The model induced on $\bZ$ is therefore {\em permutation invariant} \citep[PI, see e.g.~][]{berger2013statistical},  
meaning that for any permutation $\tau$, the distribution of $\tau(\bZ)$ under $\bfbeta^*$ is the same as the distribution of $\bZ$ under $\tau(\bfbeta^*)$, where $\tau(\bZ) := (Z_{\tau(1)},...,Z_{\tau(p)})$ is the reordering of $\bZ$ according to $\tau$. 
Here and throughout, $\bfbeta^*$ indexes the distribution of $\bZ$, not of $\tau(\bZ)$. 
Restricting attention to estimators $\widehat{\bfbeta}=\widehat{\bfbeta}(\bZ)$ of  $\bfbeta^*$, suppose now that the loss function is itself PI, i.e., $L(\tau(\bfbeta^*), \tau(\widehat{\bfbeta}))=L(\bfbeta^*, \widehat{\bfbeta})$ for any permutation $\tau$; for example, this is satisfied for squared loss $\|\widehat{\bfbeta}-\bfbeta^*\|^2$ or, under \eqref{eq:gram}, for the quadratic loss $(\widehat{\bfbeta}-\bfbeta^*)^\top \bX^\top \bX (\widehat{\bfbeta}-\bfbeta^*) = \|\widehat{\bfeta}-\bfeta^*\|^2$. 
In that case the entire problem is said to be PI, and 
the invariance principle calls to limit attention to the class $\calD^{PI}$ of all {\em PI estimators}, i.e., estimators which satisfy 
\begin{equation}
    \widehat{\bfbeta}(\tau(\bZ)) = \tau(\widehat{\bfbeta}(\bZ))\ \ \ \text{for any permutation }\tau. 
\end{equation}
Note that a penalized likelihood estimator \eqref{eq:pen-lik} for {\em any symmetric} function $\mathcal{P}_{\lambda}(\bfbeta)$, including the plain MLE, will be PI. 
Also, the oracle Bayes rule $\widehat{\bfbeta}_{ol}$ is PI, since the likelihood (model) is PI and the prior $\widetilde{\Pi}^*_p$ is exchangeable. 
In fact, the following proposition says that, for any fixed $\bfbeta^*$, the oracle Bayes rule is the {\em best} PI rule. 

\begin{proposition}\label{prop:oracle-bayes-optimality-freq}
In the Gaussian linear model with known $\psi$, suppose that $\bX$ has the form \eqref{eq:gram} and consider estimating $\bfbeta^*$ under a PI loss function. 
Then, among all PI estimators $\widehat{\bfbeta}(\bZ)$, the oracle Bayes rule $\widehat{\bfbeta}_{ol}$ is optimal, i.e., 
for any fixed $\bfbeta^*$, 
$$
\widehat{\bfbeta}_{ol} = 
\argmin_{\widehat{\bfbeta}\in \calD^{PI}}R(\bfbeta^*, \widehat{\bfbeta}). 
$$
\end{proposition}
We now return to the general case of a GLM with arbitrary $\bX$. 
While we are no longer able, in this general setup, to invoke permutation invariance considerations and establish optimality in terms of the frequentist risk as we did in the special case above, we can still show that $\widehat{\bfbeta}_{ol}$ is {\em Bayes}-optimal simultaneously over a large class of priors. 
To state our second result, for any fixed $p$-dimensional prior $\bfbeta\sim \widetilde{\Pi}$, let 
\begin{equation}\label{eq:bayes-rule-gen}
\widehat{\bfbeta}_{\widetilde{\Pi}}(\bY) 
=
\argmin_{\bfb\in \RR^p} \EE_{\widetilde{\Pi}}[L(\bfbeta, \bfb)\lvert \bY]
\end{equation}
 % \revv{DY: is expectation in the Bayes estimator over $\bar{P}$ or $\widetilde{\Pi}$?}
be the minimizer of the posterior expected squared loss under the prior $\widetilde{\Pi}$, i.e., this is the 
Bayes estimator under $\widetilde{\Pi}$. 
The following proposition essentially says that, if $\widetilde{\Pi}$ is {\em any} {\em exchangeable} prior, then the oracle Bayes rule $\widehat{\bfbeta}_{ol}$---which does not depend on $\widetilde{\Pi}$---attains smaller (no greater) Bayes risk under $\widetilde{\Pi}$ than the Bayes rule w.r.t.~$\widetilde{\Pi}$. 
This statement might seem a bit unusual at first, because we know that the Bayes rule under $\widetilde{\Pi}$ is the minimizer of the Bayes risk under $\widetilde{\Pi}$. 
This apparent discrepancy is reconciled by noting that, whereas $\widehat{\bfbeta}_{\widetilde{\Pi}}$ is a function of $\bY$ only, the oracle $\widehat{\bfbeta}_{ol}$ is a function of $\bY$ and $\bfbeta$. That is, our oracle gets to see also the {\em realized} vector $\bfbeta$ up to ordering, and therefore has an advantage even over the Bayes rule that corresponds to the correct prior. 
 \begin{proposition}\label{prop:oracle-bayes-optimality}
Let $\widetilde{\Pi}$ be any exchangeable prior on $\bfbeta$. Then 
$$
r(\widetilde{\Pi},\widehat{\bfbeta}_{ol}) 
%\EE_{\widetilde{\Pi}}L(\bfbeta, \widehat{\bfbeta}_{ol})
\leq r(\widetilde{\Pi}, \widehat{\bfbeta}_{\widetilde{\Pi}}), 
$$
where the oracle Bayes rule $\widehat{\bfbeta}_{ol}$ is essentially given by \eqref{eq:oracle-bayes}, except that, to be precise, we need to condition also on $\{\bfbeta\}$ because $\bfbeta$ is now random. Thus, formally, $\widehat{\bfbeta}_{ol}(\bY) = \argmin_{\bfb\in \RR^p} \EE_{\widetilde{\Pi}^*_p}[L(\bfbeta, \bfb)\lvert \bY, \{\bfbeta\}]$. 
\end{proposition}
The proofs for Propositions \ref{prop:oracle-bayes-optimality-freq} and \ref{prop:oracle-bayes-optimality} are provided in the Appendix. 

\begin{remark} There is a slight abuse of notation in the inequality displayed in the statement of Proposition \ref{prop:oracle-bayes-optimality} because, formally, the Bayes risk $r(\widetilde{\Pi}, \widehat{\bfbeta})$ is defined only for legal estimators, whereas $\widehat{\bfbeta}_{ol}$ depends also on the true $\bfbeta$. 
Thus, writing $\EE_{\widetilde{\Pi}}L(\bfbeta, \widehat{\bfbeta}_{ol})$ is more accurate than 
$r(\widetilde{\Pi},\widehat{\bfbeta}_{ol})$, but we chose to use the latter because it makes the statement clearer. 
\end{remark}

As a simple consequence of Proposition \ref{prop:oracle-bayes-optimality} we have 
% the following corollary, which applies to {\em any} estimator, not necessarily a Bayes estimator. 

\begin{corollary} \label{cor1}
For any estimator $\widehat{\bfbeta}  = \widehat{\bfbeta} ( \bY)$ and any exchangeable prior $\widetilde{\Pi}$,
$$
r(\widetilde{\Pi}, \widehat{\bfbeta}_{ol}) \leq r(\widetilde{\Pi},\widehat{\bfbeta}). 
% \EE_{\widetilde{\Pi}}L(\bfbeta, \widehat{\bfbeta}_{ol})\leq \EE_{\widetilde{\Pi}}L(\bfbeta, \hat{\bfbeta}).
$$
 \end{corollary}

\begin{proof}
Recalling that 
$\EE_{\widetilde{\Pi}}L(\bfbeta, \widehat{\bfbeta}_{\widetilde{\Pi}})\leq \EE_{\widetilde{\Pi}}L(\bfbeta, \hat{\bfbeta})$ by the definition of $\widehat{\bfbeta}_{\widetilde{\Pi}}$, 
this follows immediately from Proposition \ref{prop:oracle-bayes-optimality}.
\end{proof}

We now turn to explaining why we intuitively expect the methods proposed in Section \ref{sec:method} to approximate (estimate) the oracle Bayes rule. 
Our hierarchical Bayes (hBayes) estimates are  based on the posterior distribution of $\psi$ and $\boldsymbol{\beta}$ under the generative model, which can be written
\begin{equation} \label{eq:gen:pst}
f ( \psi, \bfbeta | \vec{Y}) = \int f ( \psi, \bfbeta, \boldsymbol{\phi} | \vec{Y}) d \boldsymbol{\phi} = 
\int f ( \psi, \bfbeta |  \boldsymbol{\phi},  \vec{Y}) f ( \boldsymbol{\phi} | \vec{Y}) d \boldsymbol{\phi}. 
\end{equation}
Recall that in the generative model,  $\boldsymbol{\phi}$ is the random parameter vector that specifies $\Pi$, 
the marginal distribution of the coefficients $\beta_j$.
Thus, $f ( \psi, \bfbeta |  \boldsymbol{\phi},  \vec{Y})$ is the posterior distribution of the parameters $\bfbeta$ and $\psi$ of the model \eqref{eq:glm}, under the prior specified by \eqref{eq:prior-psi} and \eqref{eq:prior-iid} for a {\em fixed} distribution $\Pi$. 
Setting the FPT model hyper-parameters $( \alpha_{l, i}, \beta_{l, i})$ to $1$ makes the prior (marginal) distribution of $\beta_j$ 
the uniform density on $[a_{min}, a_{max}]$, reflecting a high degree of uncertainty regarding the distribution of the components of $\bfbeta$. 
In the very special case where $n=p$ and $\bX = \boldsymbol{I}_p$, \eqref{eq:glm} reduces to a sequence model (with ``free", i.e., unrelated,  parameters $\eta_j = \beta_j$), and our method resembles the hierarchical constructs in \citet{antoniak1974mixtures}. 
In the sequence model, putting a `nonparametric' noninformative prior on $\Pi$ is a fully Bayes alternative to (nonparametric) empirical Bayes strategies, which, in turn,  
%is generally known to 
pursue consistent estimation of 
%exactly 
the empirical distribution $\widetilde{\Pi}^*_p$ of the $\beta_j$'s \citep[see, e.g.][]{zhang2003compound, brown2009nonparametric}, in a model where $\bfbeta$ is fixed. 
% \color{magenta}Gosia: I have not found this statement in the zhang paper. Maybe Brown and Greenstein (2009) would be a better reference ? You mean consistent estimator, or optimal estimator ? I wrote efficient  (based on Brown and Greenstein), but I am not sure.\color{black}
Our hierarchical model in Section \ref{sec:method} extends these ideas to regression models (GLMs), namely, by carrying over the noninformative Polya tree prior on the coefficients of the more genera model \eqref{eq:glm}, we still expect the posterior of $\Pi$ to approximate $\widetilde{\Pi}^*_p$, the empirical distribution of the (fixed) $\beta_j$'s. 
It is worth remarking here that the particular choice of a Polya tree prior is just one option we found to work well in our experiments, but other `noninformative' choices such as distributions based on Dirichlet priors, could also be considered. 
In our simulation studies we will show that the posterior mean of $\Pi$ under the generative model does indeed provide a good approximation of $\widetilde{\Pi}^*_p$. 
Referring to the representation \eqref{eq:gen:pst}, this implies that $f ( \boldsymbol{\phi} | \vec{Y})$ assigns large weights 
to $\boldsymbol{\phi}$ corresponding to distributions $\Pi$ which are similar to $\widetilde{\Pi}^*_p$. 
In turn, according to \eqref{eq:gen:pst}, the posterior distribution of $\psi$ and $\boldsymbol{\beta}$, and the corresponding Bayes estimates of the $\beta_j$'s 
under the generative model,  will be similar to the posterior distributions and Bayes rules under the prior \eqref{eq:prior-iid} with $\Pi$ replaced by $\widetilde{\Pi}^*_p$. 

% In the next section we refer to the latter as an ``oracle Bayes" estimator.  
% \sout{We call these oracle distributions and oracle Bayes rules and we will formally define them
% in the the next section.}

\section{Simulations}
\label{sec:simulations}

We turn to a simulation study for demonstrating the utility of our methods. 
We focus on a logistic regression model and use the simulation setup of \cite{sur2019modern}, where $\bX$ consists of  $n = 4000$ rows and $p = 800$ columns of  i.i.d. $\calN(0, 1/n)$ entries. 
Three different configurations for the coefficient vector $\bfbeta$ are considered: (I) $\bfbeta$ has $100$ replicates  of $-10$, $100$ replicates of $+10$, and $600$ zeros; 
(II) $\bfbeta$ consists of $800$ i.i.d. $\calN(3, 16)$ realizations; 
(III) $\bfbeta$ consists of $400$ i.i.d.~ $\calN(7, 1)$ realizations, and $400$ zeros. 
For each of the experiments (I)-(III) we ran $30$ Monte Carlo rounds,  generating
\begin{equation}\label{eq:logistic}
Y_i\sim \textit{Bernoulli} (q_i), \ \ \ \ \ q_i = 1/(1+\exp(-\bX_i^\top \bfbeta)),
\end{equation}
and holding $\bX$ and $\bfbeta$ fixed through the $30$ simulation runs. 
We calculated the {root mean square error (RMSE)}, the average over the 30 runs of  
$\sqrt{\sum_{j = 1}^{p} ( \beta_j - \hat{\beta}_j)^2 / p}$, for six estimators of the vector $\bfbeta$: 
the maximum likelihood estimator (``MLE"), $\hat{\bfbeta}_{MLE}$; the bias-corrected maximum likelihood estimator (``adj-MLE") of \citet[][]{sur2019modern}, which in all cases (I)-(III) is $\hat{\bfbeta}_{MLE} / 1.499$; 
an $L_2$-penalized estimator (``Ridge") and an $L_1$-penalized estimator (``LASSO"), implemented using the  \texttt{cv.glmnet} function from the \texttt{glmnet} package \cite{glmnet_friedman} with default specifications; and 
the proposed hierarchical Bayes estimator (``hBayes") using a 6-level Polya tree, with $a_{min}  =  \min (-24, \hat{\bfbeta}_{MLE} - 0.5)$ and $a_{max} =  \max (24, \hat{\bfbeta}_{MLE} + 0.5)$, divided evenly into $64$ subintervals, with $500$ iterations of the Gibbs sampler using the first $100$ iterations as burn-in.
For each implementation of the hBayes approach we generally set $[a_{min}, a_{max}]$ to be slightly larger than range of the components of  $\hat{\bfbeta}_{MLE}$, to ensure large overlap of the support of $\Pi (  \beta |   \vec{\pi}_L )$ 
% $\pibar (  \beta |   \vec{\pi}_L )$ 
for each simulation run in each experiment (I)-(III).
As a reference, we also computed the oracle Bayes estimator (``oBayes'') given in Equation \eqref{eq:oracle-bayes}. 
Performance of the Oracle Bayes estimates was evaluated by running $500$ iterations of a permutation Gibbs sampler, 
described in Appendix 
%B.1, 
\ref{sec:sample_oracle}, 
using the first $100$ iterations as burn-in.

Table \ref{tab:logistic} reports the RMSE for the five estimators and the oracle in each of the three experiments (I)-(III). 
%In all simulation scenarios, the (estimated) RMSE for the Lasso and Ridge shrinkage estimators is smaller than the RMSE for the adjusted MLE, which is still considerably smaller than the RMSE for the MLE.
In all three experiments the RMSE of hBayes wsa considerably 
smaller compared to Lasso and Ridge, and only slightly larger than that of the oracle. 
The approximation is particularly good in Experiment (II), where the distribution of the parameter vector is relatively easy to estimate. 
In Experiment (I), the Lasso estimates yield smaller RMSE than Ridge estimates, while in Experiments (II) and (III) Lasso had larger estimation error  than Ridge. 
%In Scenario 2, where the distribution of the parameter vector is relatively easy to estimate using our hierarchical Bayes approach,  the difference in RMSE between the hBayes and oBayes estimates was very small. 
%Figure \ref{fig:oracle-Bayes}, included in the Appendix, displays the values of $\widehat{RMSE}$ for the 30 runs of each scenario. 
%t shows that the hBayes and oBayes methods yield smaller estimated RMSE for each run of the simulation in all three scenarios. 
%The small run-to-run variability of the estimated RMSE, 
It also appears that hBayes and oBayes yield parameter estimates that are considerably more stable than Ridge and Lasso, as indicated by the smaller standard errors in Table \ref{tab:logistic}. 

%in each realization of the simulation. 

\begin{table*}[!ht]
\centering
\small
\begin{tabular}{l|lllllllll}
&  MLE & oBayes & hBayes & LASSO &  Ridge & adj-MLE \\     \hline
Experiment (I) & $5.11 \  (0.06)$  &  $1.86 \ (0.03) $  &  $\mathbf{1.97} \ (0.02)$ &  $2.53  \ (0.04)$   & $ 2.85 \ (0.06)$  & $ 3.05 \ (0.03)$  \\   \hline 
Experiment (II) & $5.21 \ (0.06)$ &  $2.36  \ (0.01)$  & $ \mathbf{2.38} \ (0.01)$   &  $3.09 \ (0.10)$ &   $2.91  \ (0.05)$  & $ 3.12 \ (0.03)$ \\   \hline 
Experiment (III) & $5.25 \ (0.07)$ &  $ 2.02  \ (0.01)$ & $\mathbf{2.10}  \ (0.01)$   &  $  3.06 \ (0.12)$  &  $ 2.90  \ (0.07)$ & $ 3.11 \ (0.03)$  \\   \hline 
  \end{tabular}
\caption{Root mean square error (RMSE) in the three simulated examples.
Numbers are averages of the RMSE from $30$ simulation runs of each simulated example. 
Parentheses show standard errors for these averages}  
\label{tab:logistic}
\end{table*}

\begin{figure*}
\centering
\includegraphics[width=.95\textwidth, height=.325\textwidth]{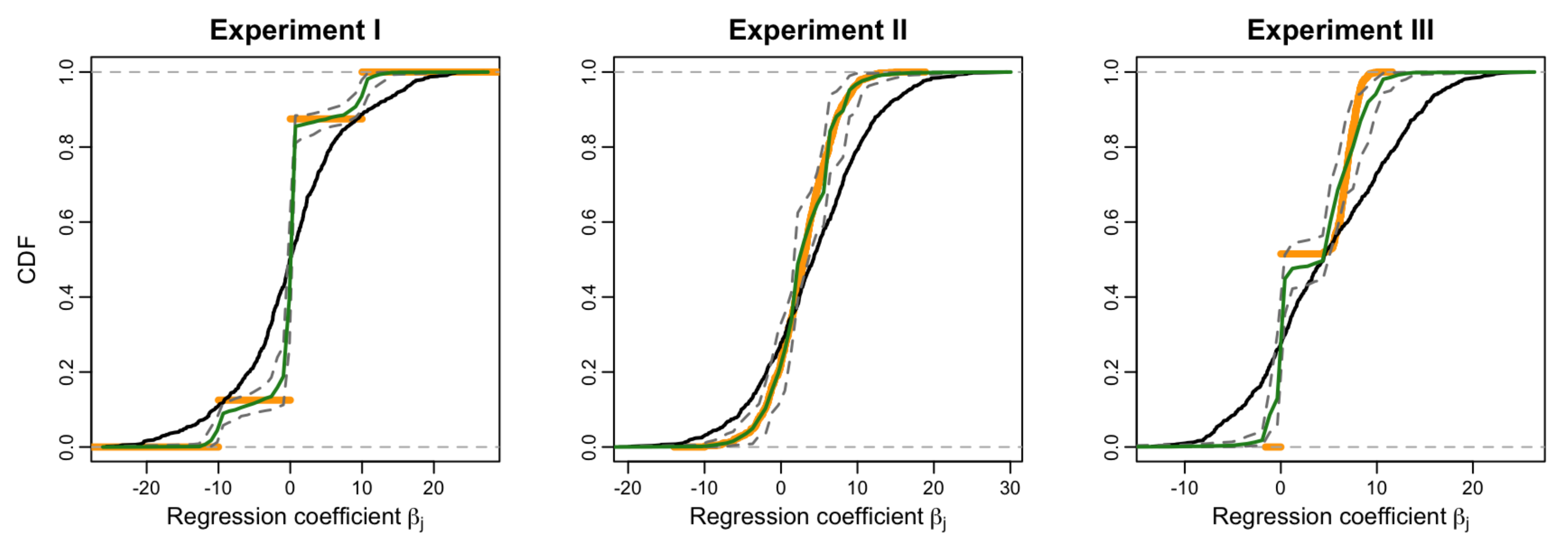}
\caption{
Empirical CDF of the true coordinates $\beta_j$ of $\bfbeta$ (orange), and estimates thereof, for the simulation experiments (I)-(III). 
Black curve is empirical CDF of the $p$ components of $\hat{\bfbeta}_{MLE}$. 
Solid green curve is the (estimated) posterior median of the CDF of $\Pi$ under the generative model for the proposed method. 
Dashed green lines mark $0.025$ and $0.975$ (estimated) quantiles of this posterior. 
}
\label{fig:ecdf}
\end{figure*}

Figure \ref{fig:ecdf} presents, for each experiment (I)-(III), the empirical cumulative distribution function (CDF) of the coordinates $\beta_j$ of the {\em true} vector $\bfbeta$, 
along with estimates of this (true) empirical CDF corresponding to the MLE and the proposed method, see caption for details. 
In all three scenarios our hierarchical Bayes method was able to recover the overall shape of the true empirical CDF, whereas this shape is undetectable from the empirical CDF of the noisy maximum likelihood estimates. 
Our approach produces smoothed distribution estimates that are shrunk toward the uniform distribution. 
This can be seen more clearly in Scenarios 1 and 3, where the distribution of the $\beta_j$'s has point mass. 
% ; for scenario 2, 
% in which $\beta_j$ were drawn from a continuous distribution, our method yielded particularly good estimates of the true empirical CDF. 

\section{Unraveling polygenic inheritance}
\label{sec:real-data}

We now apply our hierarchical Bayes approach 
%to real data, and use our methods 
to gain insight into the genetic architecture of polygenically inherited traits. 
In this context the explanatory variables are appropriately coded genotypes of genetic markers and the vector of regression coefficients represents the influence of specific genomic regions on the trait.

 Many genetic studies point out that genetically inherited traits are often influenced  by many genes with small effects, distributed over the whole genome \citep{P_08,F_10, F_11,T_12,VH_96, Magnus1}. 
As discussed in, e.g., \cite{Wallin_22}, analyzing the respective genetic data with classical ``sparse"  regression models leads to highly unsatisfactory results. 
Instead, geneticists often use mixed linear models, where the polygenic effect is represented by one random effect, meant to capture the effect of all polygenes (see e.g., \cite{EMMAX}), or by many small random effects at all markers (\cite{Piepho_09, Endelman_11}). 
In the latter case the estimation of individual effects is often performed using ridge regression, which yields an empirical version of the BLUP when the genetic effects arise from the Normal distribution and the tuning parameter is adjusted according to the ratio of the variance of this distribution and the variance of the noise term. 
In \cite{Wallin_22} the classical mixed model is further extended by allowing a nonzero mean in the random effect, which is suitable when the investigated population is an admixture of populations subject to different selection pressures.

All methods mentioned above assume that the polygenic effects come from a Normal distribution. 
As shown in the real data analysis below, this assumption can be grossly inadequate. Thus, unraveling the inheritance of polygenic traits  is an interesting case for our nonparametric Bayes approach. 
In the following section we report the results of a simulation study and a real data analysis to illustrate the advantages of our method in analyzing such genetic data. 
\subsection{Simulation study}\label{subsec:real-data-simulation}
Our study follows the design of the simulation study for the experimental backcross design from \cite{Wallin_22}. Thus, we simulated data for $n=400$ individuals from the backcross, where the marker genotypes can take only two values, $X_{ij}\in\{1,-1\}$, which coincide with the ancestry indictors (i.e. parental line indicators) at given loci. 
We simulated 10 chromosomes, each of the length of 150cM (centiMorgan), with markers spaced every 1cM. This means that for the consecutive markers on the same chromosome $P(X_{ij}=X_{i(j+1)})\approx 0.99$, while the markers on different chromosomes are independent. 
Following \cite{Wallin_22}, the trait values are generated as 
\begin{equation}\label{polym}
\bY= \bX\bgamma + \bX_m\tilde \bfbeta +\bepsilon,
\end{equation}
where  $\bX$ is the $400\times 1500$ incidence matrix with all marker genotypes, 
$\bepsilon \sim \mathcal{N}_{400}\left(\boldsymbol{0},0.1 \boldsymbol{I}\right)$, and $\bX_m$ is the $400\times 4$ matrix, whose first column consists of all ones (to model the intercept term) and the remaining three columns form a  subset of $\bX$ containing  genotypes of markers strongly associated with the trait, $\tilde \bfbeta=(\tilde \beta_0,\ldots,\tilde \beta_{3})\in \mathbb{R}^4$. 
The elements of the polygenic random effects vector $\bgamma=(\gamma_1,\ldots,\gamma_{1500})$ are i.i.d.~random variables from a generalized Laplace distribution, where the Normal mean-variance mixture is of the form 
$$
\gamma_j  =  \mu  + \tau \left( \xi (V_i -1) +  \sqrt{V_i} Z_i\right), 
$$

with $V_i \sim \Gamma(\nu,\nu) $ and $Z_i \sim \mathcal{N}(0,1)$. 
The parameter $\mu$ represents the expected value of $\gamma_j$, $\xi$ controls the asymmetry of the distribution, and $\nu$ the shape of the distribution. 
In our simulation we set $(\mu,\xi,\tau,\nu)=(-0.01,-2,0.05,0.75)$, which generates a spiked, strongly asymmetric distribution with exponential tails. 
We set
$\bX_m = [1_{400}, X_{300}, X_{750}, X_{1200}]$ and $\bfbeta=(0,0.2,0.2,-0.2)$,  s.t.~the first two signals are in the opposite direction of the polygenic effect, and the third  is in the same direction.

We analyze our simulated data sets with the  proposed hierarchical Bayes approach and with some classical and modern methods for the analysis of high-dimensional regression models. 
% We include the group of methods aimed at discovering the sparse signals like LASSO or the Bayesian methods using the classical spike and slab priors, and the group of methods using the Normal or the mixtures of Normal priors, more suitable for the situation when the signal is dense. 
Among methods targeting sparse signals, we included LASSO \citep{tibshirani1996regression} with the tuning parameter selected by cross-validation, and four Bayesian variable selection methods: 
the Sum-of-single-effects method (SuSiE) by \cite{wang2020simple};  the {\it varbvsmix} algorithm of \cite{carbonetto2017varbvs}, as implemented in the {\it varbvs} package in \texttt{R} (VARBVSMIX); Spike-and-Slab LASSO (SSL) of \cite{rovckova2018spike}; and the Expectation-Maximization Variable Selection (EMVS) method of \cite{rovckova2014emvs}. 
Among methods targeting dense signals, we included Ridge Regression \citep{ridge} with the tuning parameter selected by cross-validation; Bayesian Multiple Regression with Adaptive Shrinkage (Mr.~ASH) by \cite{kim2022flexible}; and the method of \cite{Wallin_22} based on the mixed regression model (\ref{polym}). 

\begin{table*}[!ht]
\small
\centering
\caption{Estimates of MRNE obtained by averaging RNE over 200  simulation runs}
\label{Table:RMSE_ICML}
\vspace{0.5em}
% AW: I changed "\begin{tabular}{r|rrrrrr||rrrr|r|}" 
\begin{tabular}{|r|rrrrrrrrrr|r|}  \hline & lasso &susie &vbsmix& ssl & emvs& horseshoe &ridge &mr.ash &  mix & hBayes & oracle
	 \\   \hline
	   $\hat \bfbeta $ & 0.89 &  3.10 & 1.10&3.81 &0.94 & 0.95 & 0.96 & 0.95& 0.77 & 0.76  & 0.74 \\  
	    $\hat \bfbeta_s $ & 0.75 &4.07 & 0.89 &3.40 &0.61& 0.63&  0.76 & 0.72& 0.48 & 0.46 & 0.45 \\  
	    $ \hat \bfeta $ & 0.07 & 0.31 & 0.11 & 0.24 & 0.03& 0.03 & 0.02 & 0.07& 0.02 & 0.02  & 0.02 \\
	    \hline
\end{tabular}
\end{table*}

\begin{table*}[!ht]
	\centering
\caption{Estimated MRNE for the large fixed effects}
\label{Table:sim_qtl}
\vspace{0.5em}
\begin{tabular}{|rrrr|}  \hline & $ {\bf \beta}_1$ & $ {\bf \beta}_2$ & $ {\bf \beta}_3$ \\   \hline
	lasso & 0.11 & 0.11 & 0.09 \\   ridge & 0.19 & 0.19 & 0.18 \\   mix & 0.03 & 0.06 & 0.03 \\   hBayes & 0.05 & 0.07 & 0.04 \\   
	mr ash & 0.19 & 0.19 & 0.17 \\
	EMVS & 0.12 & 0.13 & 0.11 \\ 
 SSlasso & 0.21 & 0.22 & 0.20 \\ 
 varbvsmix & 0.20 & 0.20 & 0.14 \\	
oracle & 0 & 0 & 0 \\	
	 \hline\end{tabular}
  \begin{tabular}{|rrrr|}  \hline smoothed & $ {\bf \beta}_1$ & $ {\bf \beta}_2$ & $ {\bf \beta}_3$ \\   \hline
lasso & 0.07 & 0.07 & 0.05 \\   ridge & 0.10 & 0.10 & 0.12 \\   mix & 0.02 & 0.02 & 0.02 \\   hBayes & 0.02 & 0.02 & 0.02 \\
	mr ash & 0.18 & 0.18 & 0.02 \\  
	EMVS & 0.10 & 0.11 & 0.10 \\
	SSlasso & 0.21 & 0.22 & 0.19 \\ 
	varbvsmix & 0.21 & 0.21 & 0.07 \\ 
oracle & 0.02& 0.02 & 0.02 \\	
	\hline 
\end{tabular}
\end{table*}

%  \begin{figure*}
% 	\centering
% \includegraphics[width=0.3\textwidth, height=3.8cm]{figs/beta_violon.pdf}
% \includegraphics[width=0.3\textwidth, height=3.8cm]{figs/beta_smooth_violon.pdf}
% \includegraphics[width=0.3\textwidth, height=3.8cm]{figs/xbeta_violon.pdf}
% 	\caption{Empirical distribution of the relative norm of the error (RNE)} on 200 simulation runs. 
% 		}
%         \label{fig:sim_rmse_text}
% \end{figure*}

\begin{figure}[h]
    \centering
\includegraphics[width=0.3\textwidth, height=3.8cm]{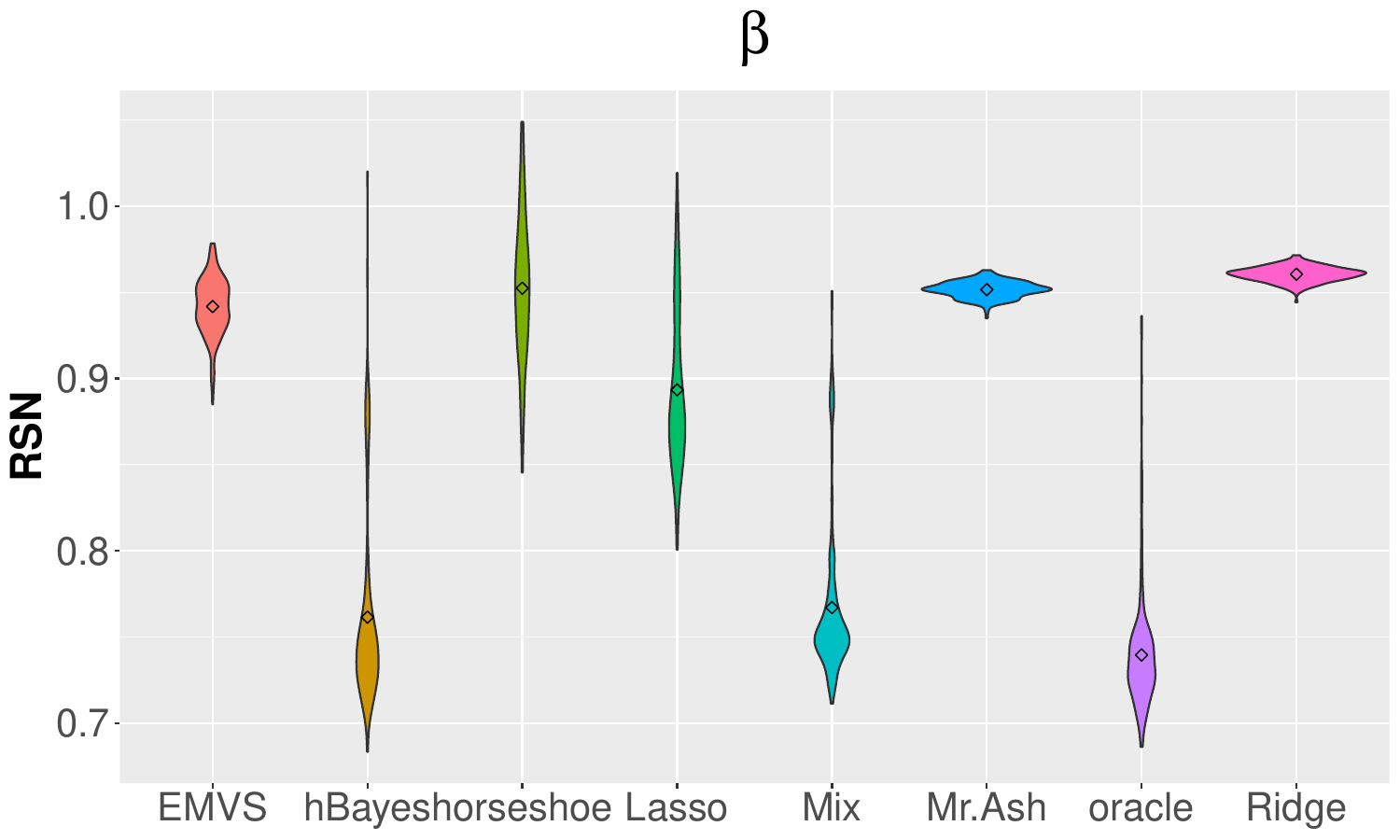}
 \includegraphics[width=0.3\textwidth, height=3.8cm]{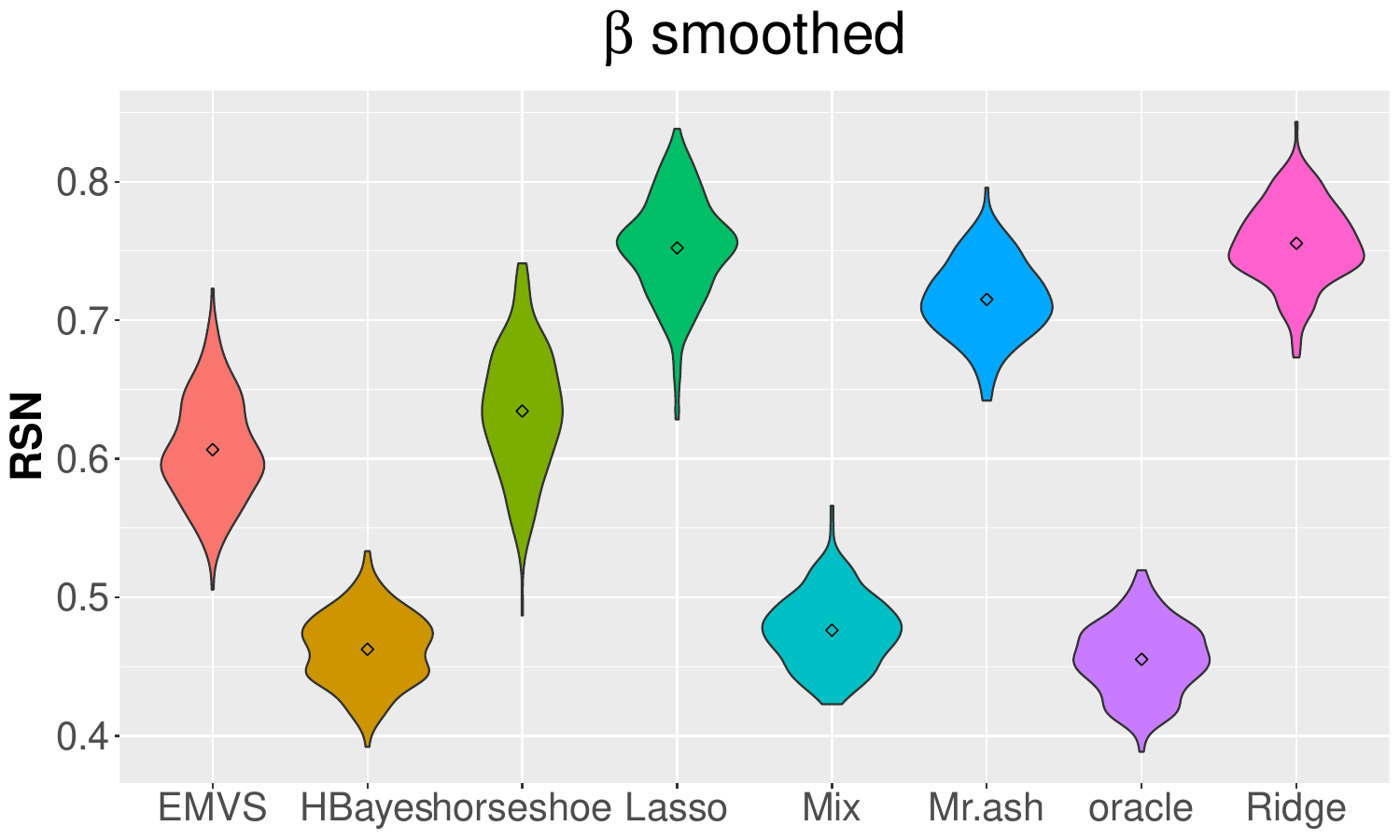}
\includegraphics[width=0.3\textwidth, height=3.8cm]{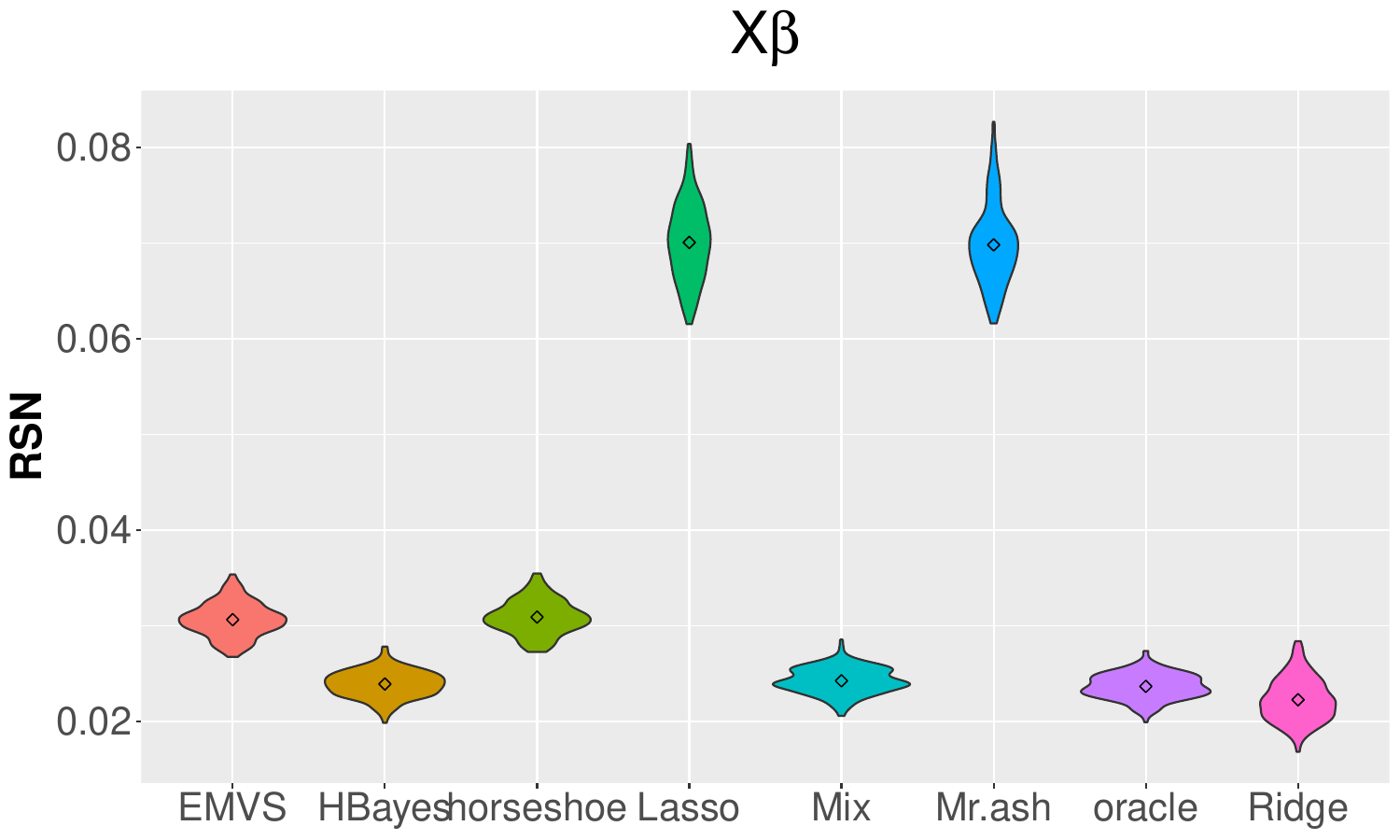}
    \caption{Empirical distribution of relative norm of the error (RNE) on 200 simulation runs}
\label{fig:sim_rmse_text}
\end{figure}

In Tables \ref{Table:RMSE_ICML} and \ref{Table:sim_qtl}, we  provide the mean relative norm of the error (MRNE) in estimation, 
$$
MRNE(\hat \beta)=\EE\left(\frac{||\hat \bfbeta - \bfbeta||}{|| \bfbeta - \bar{\bfbeta} ||} \right), 
$$
where $\bar{\bfbeta}=\frac{1}{p}\sum_{j=1}^p \beta_j$. 
Empirical distributions of the relative norm of the error (RNE) are visualized in Figure \ref{fig:sim_rmse_text}. 
Due to the strong correlation between neighboring markers, it is quite difficult to appropriately estimate individual genetic effects. 
As an alternative approach, we propose to estimate a smoothed version of $\boldsymbol{\beta}$, denoted $\boldsymbol{\beta}_s$, by a correspondingly smoothed version of $\hat{\boldsymbol{\beta}}_s$. 
Thus, we define 
\[
{MRNE}(\hat{\boldsymbol{\beta}}_s) = \mathbb{E} \left( \frac{\lVert \hat{\boldsymbol{\beta}}_s - \boldsymbol{\beta}_s \rVert}{\lVert \boldsymbol{\beta}_s - \bar{\boldsymbol{\beta}}_s \rVert} \right),
\]
where $\boldsymbol{\beta}_s$ and $\hat{\boldsymbol{\beta}}_s$ are obtained by averaging $\boldsymbol{\beta}$ and $\hat{\boldsymbol{\beta}}$, respectively, over $\pm$5 cM windows within the chromosome boundaries.
Table \ref{Table:RMSE_ICML} and Figure \ref{fig:sim_rmse_text} display the results for the entire vector $\bfbeta$, while Table \ref{Table:sim_qtl} shows the accuracy in estimating the three larger fixed effects, $\beta_1$, $\beta_2$ and $\beta_3$. 
Table \ref{Table:RMSE_ICML} and Figure \ref{fig:sim_rmse_text} include also MRNE of the error in prediction, i.e., in estimating $\bfeta = \bX\bfbeta$. 
All MRNEs are estimated based on 200 independent replicates of the entire experiment, where each uses an independent draw of the design matrix $\bX$, the vector of polygenic effects $\bgamma$ and the vector $\bepsilon$ of error terms. 
As shown in Table \ref{Table:RMSE_ICML}, two methods targeting sparse signals, SuSiE and SSL, perform poorly on our example. 

\begin{figure}%[!ht]
	\centering
\includegraphics[width=0.3\linewidth]{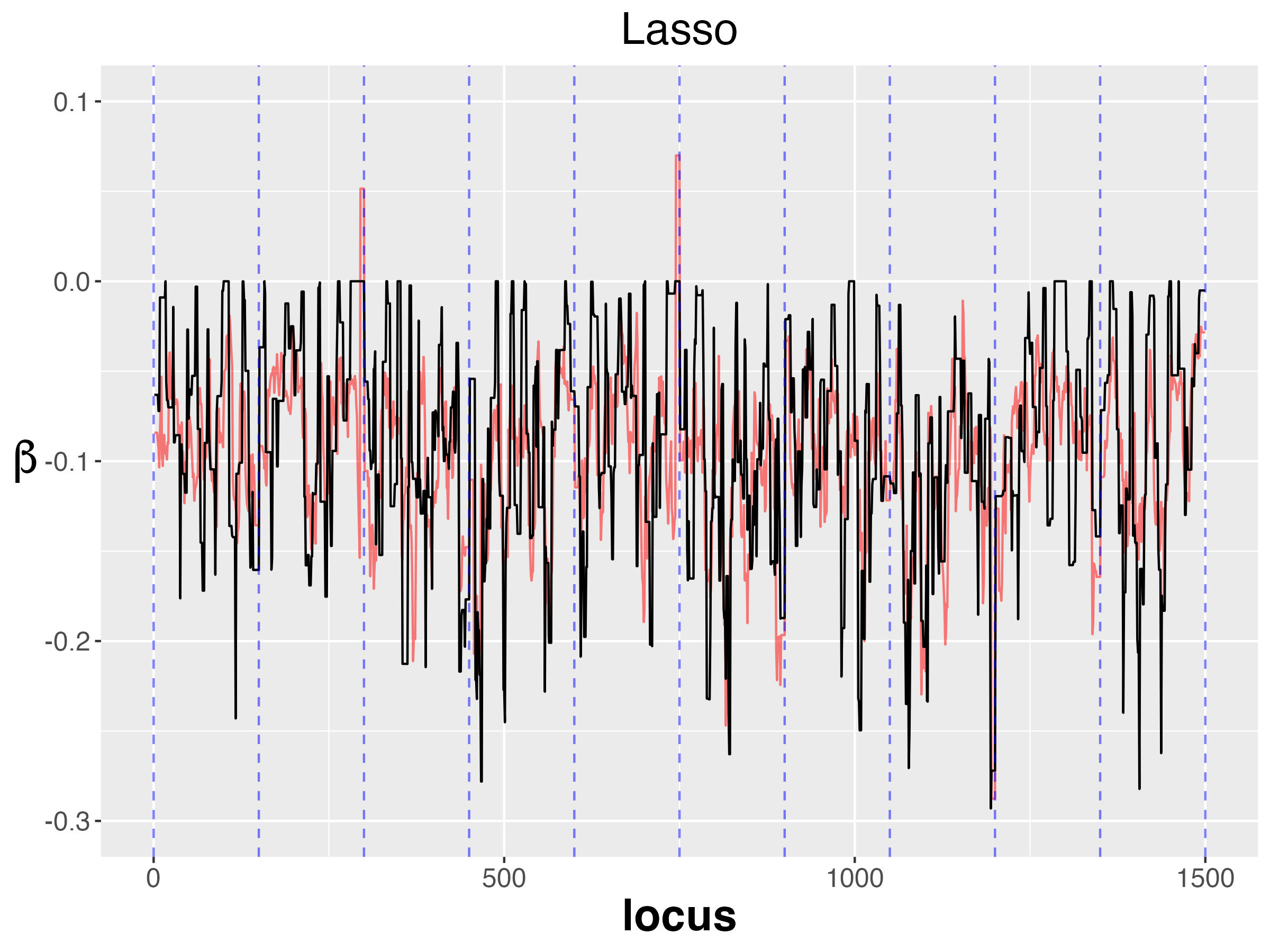}
\includegraphics[width=0.3\linewidth]{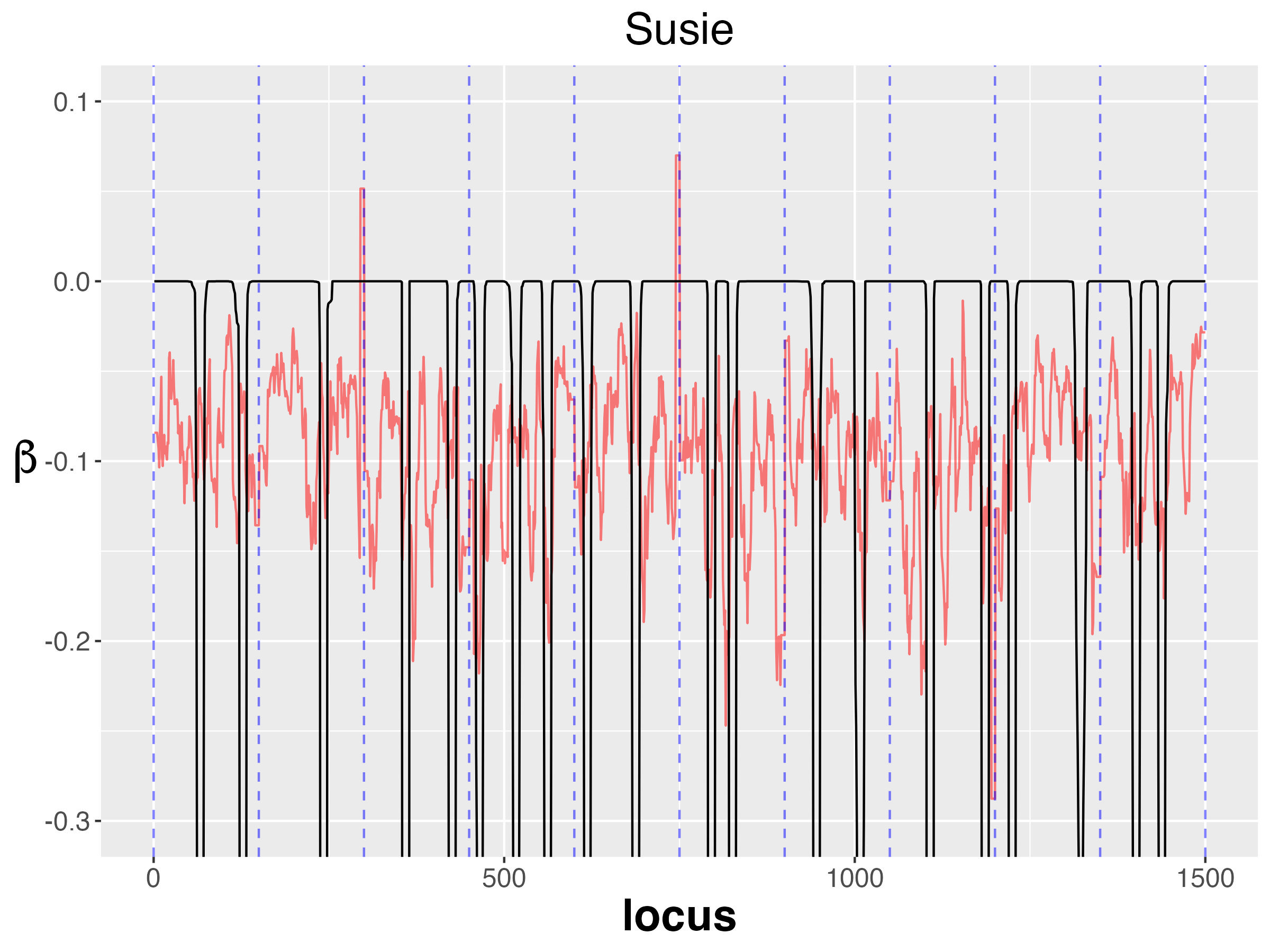}
\includegraphics[width=0.3\linewidth]{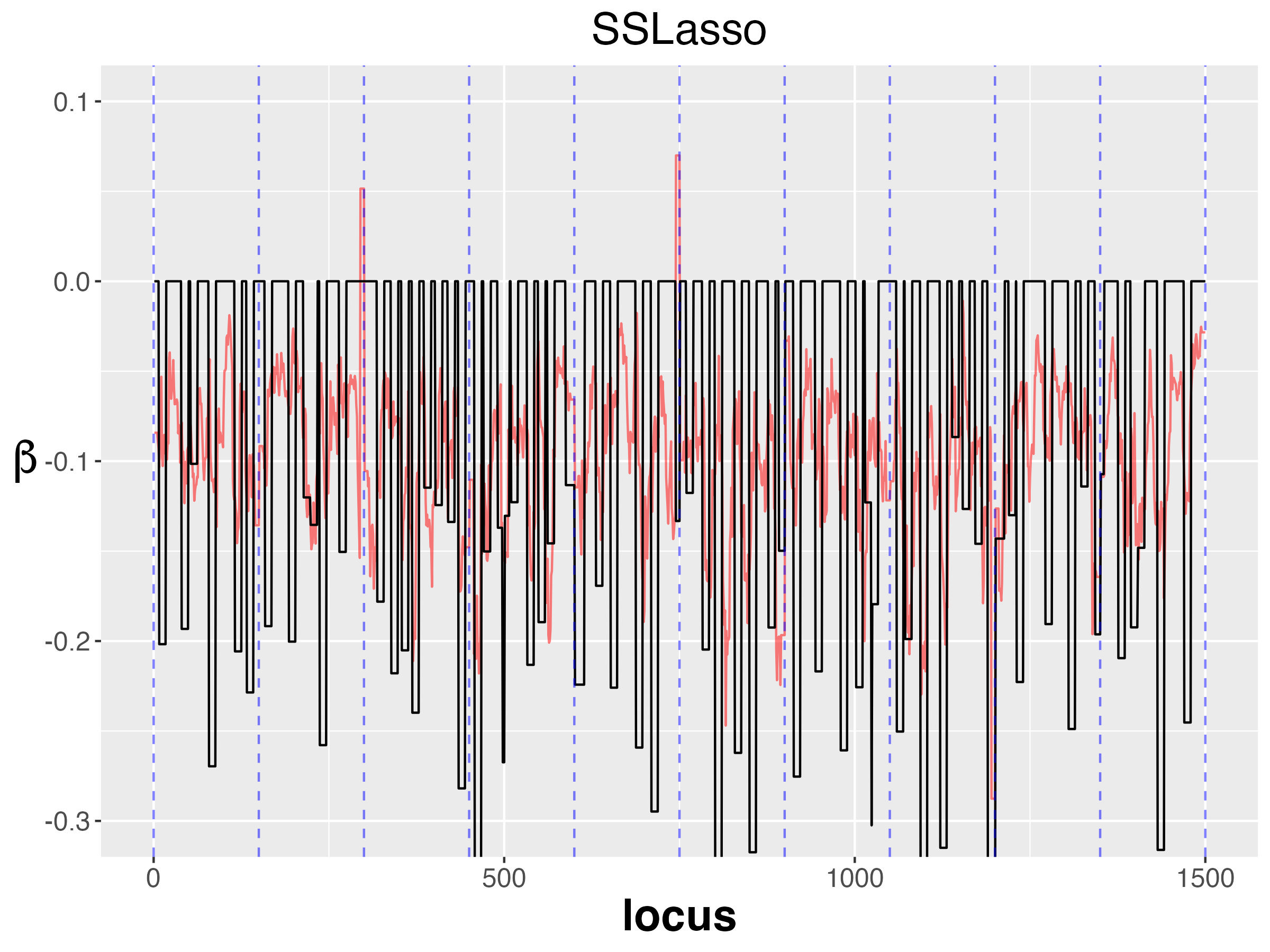}
\includegraphics[width=0.3\linewidth]{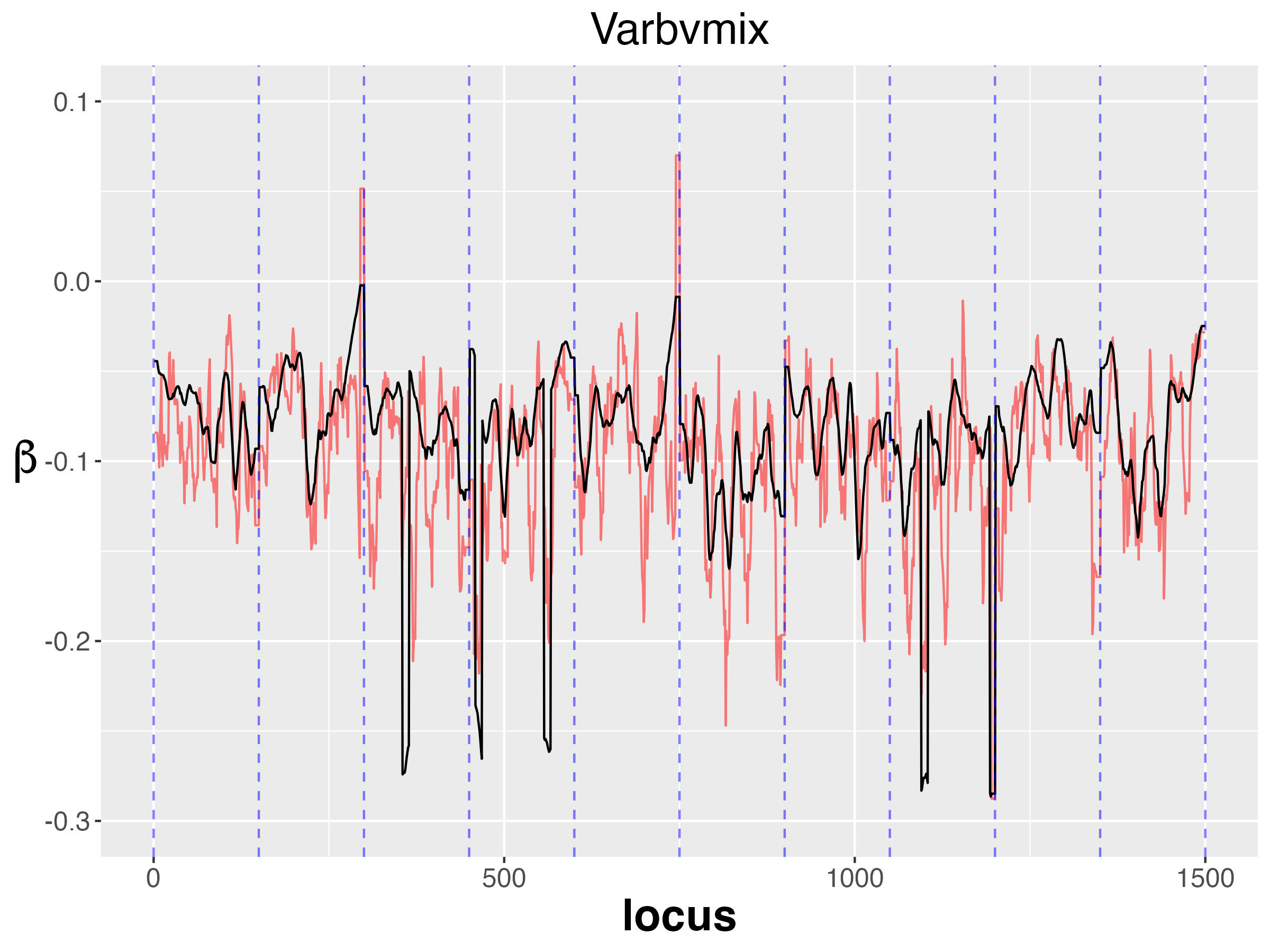}
\includegraphics[width=0.3\linewidth]{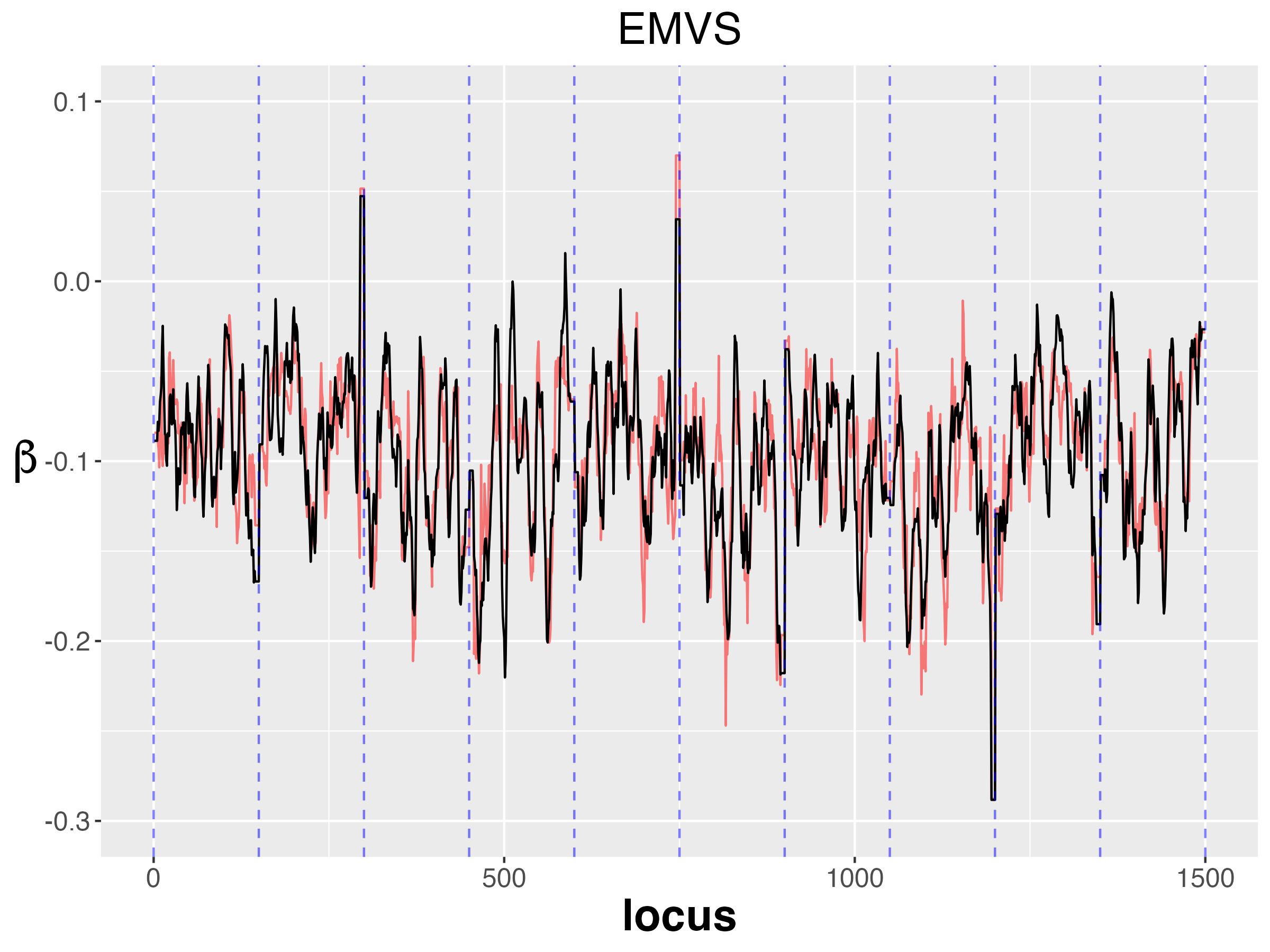}
\caption{Results of the analysis of one simulated data by the methods aimed for the selection of important variables under the sparsity assumption. Going left to right, up to down we have: first plot is the cross-validated LASSO, second is the {\it SuSiE} algorithm of \cite{wang2020simple}, followed by the spike and slab lasso of \cite{rovckova2018spike}, and {\it varbvsmix} algorithm of \cite{carbonetto2017varbvs} and {\it EMVS} algorithm of \cite{rovckova2014emvs}. Red lines mark the true genetic effects and the black lines their estimates}
\label{fig:dense_simulation_1}
\end{figure}

Figure \ref{fig:dense_simulation_1} shows that these methods result in many large false discoveries in the direction of the summary polygenic effect, and fail to locate two signals of the opposite sign. 
These observations align with the findings of \cite{Wallin_22}, which highlight the shortcomings of ``sparse" signal methods when applied to traits with a polygenic component. 
Two other spike and slab methods, \textit{varbsmix} and \textit{EMVS}, perform substantially better, with \textit{EMVS} adapting particularly well and performing similarly to methods targeted at dense signals. 
It is also interesting to note the relatively good performance of cross-validation LASSO.

\begin{figure}%[!ht]
\includegraphics[height=.275\textwidth, width=0.5\linewidth]{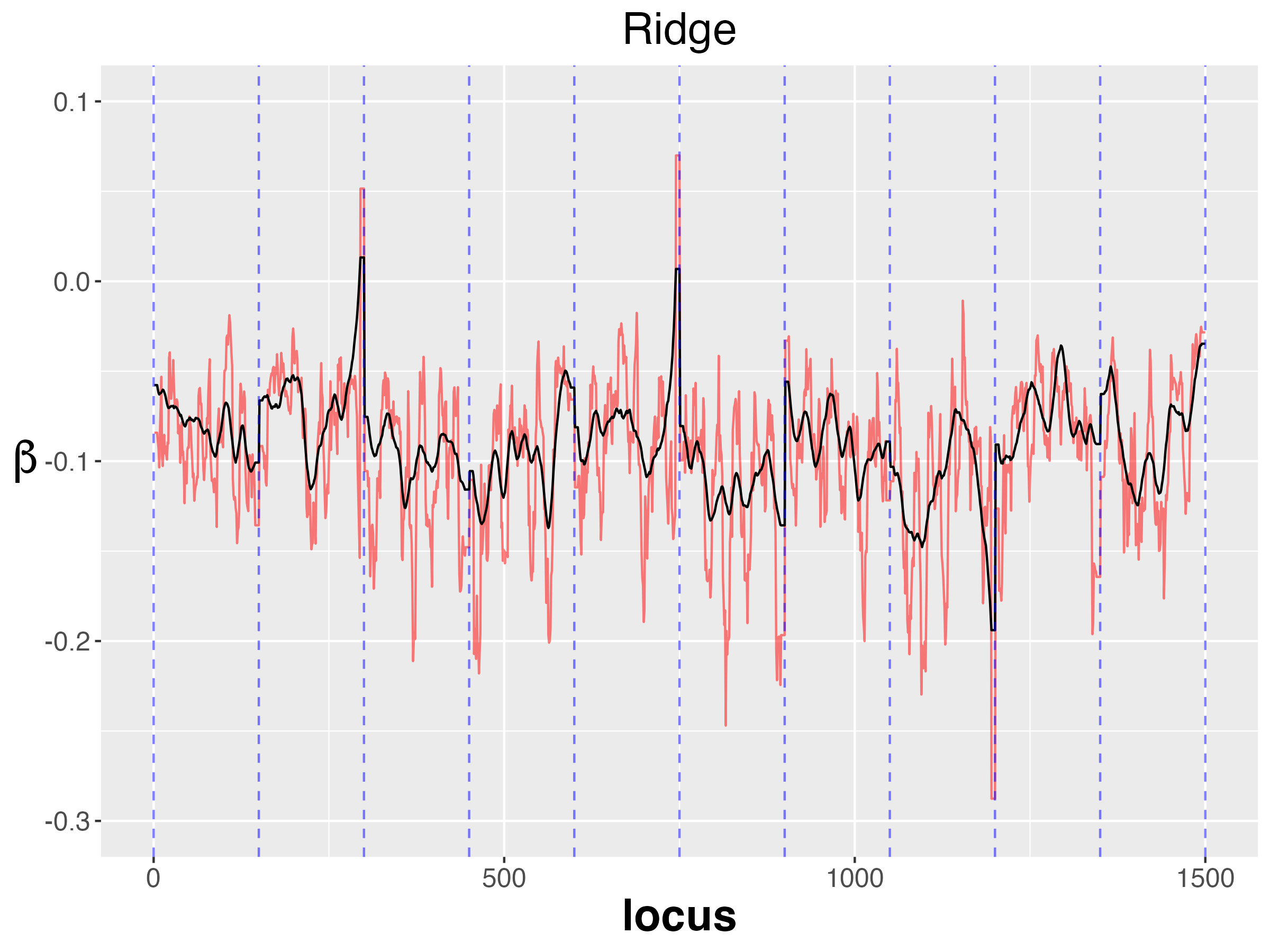}
\includegraphics[height=.275\textwidth, width=0.5\linewidth]{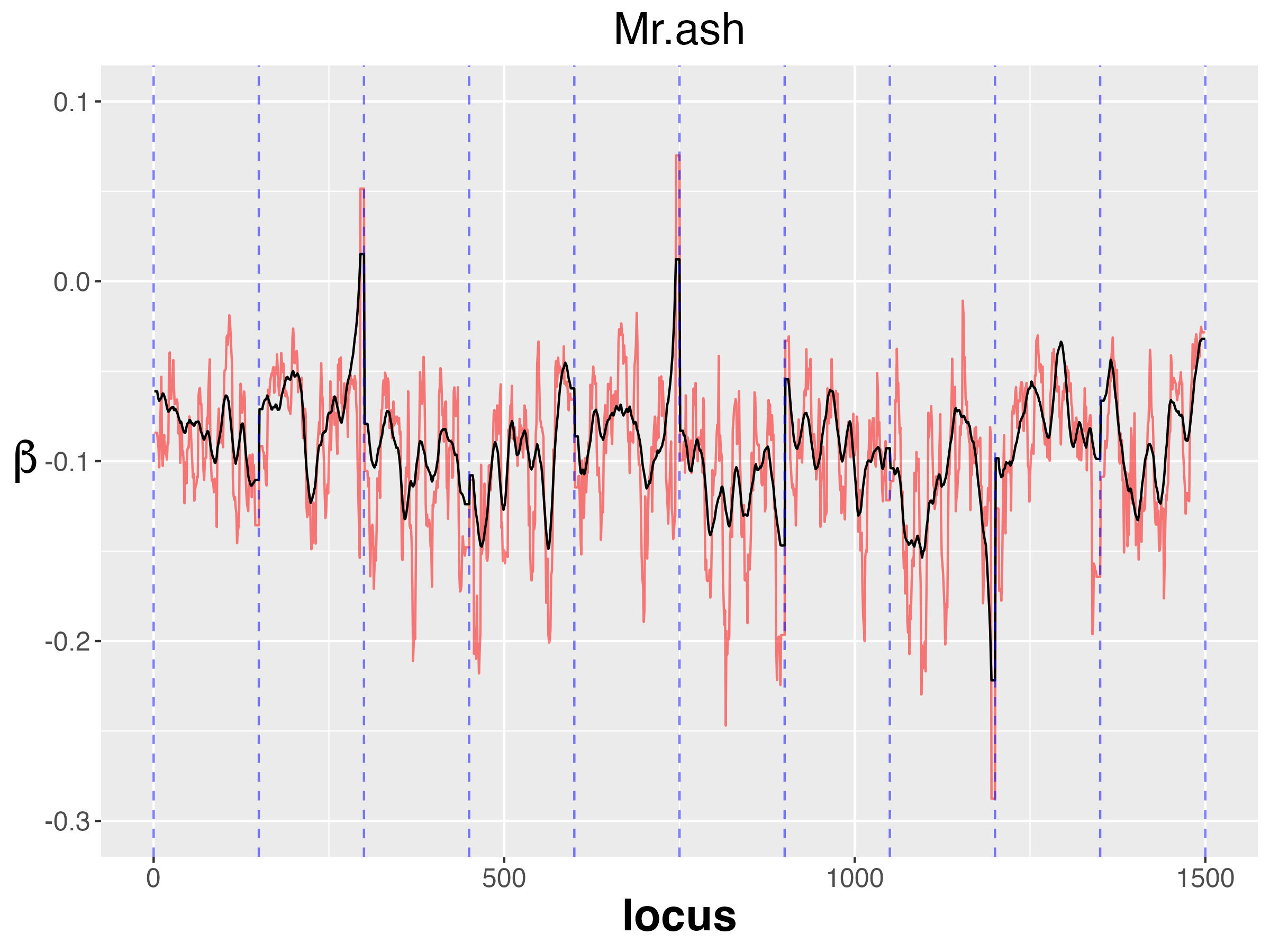}
\includegraphics[height=.275\textwidth, width=0.5\linewidth]{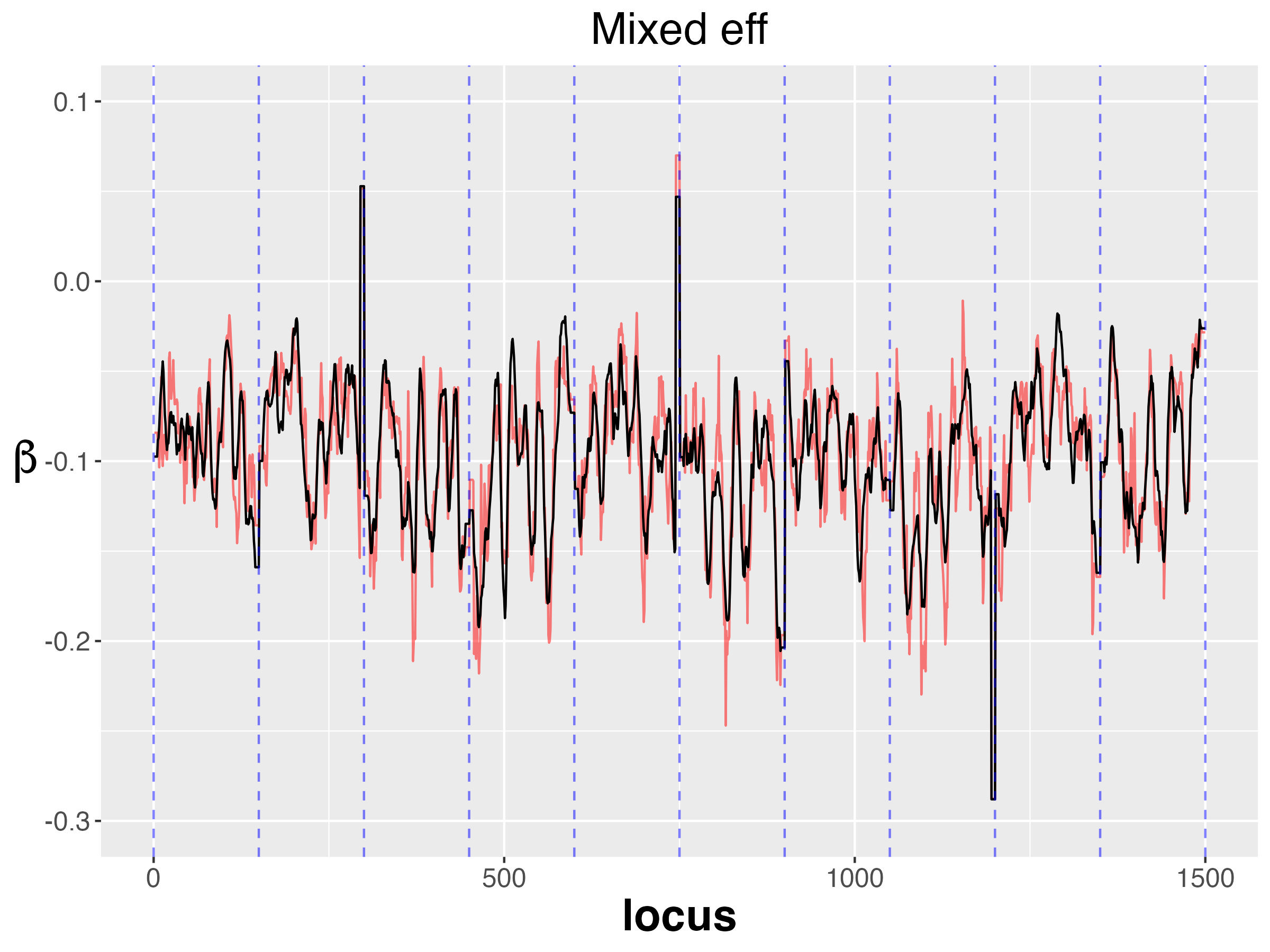}
\includegraphics[height=.275\textwidth, width=0.5\linewidth]{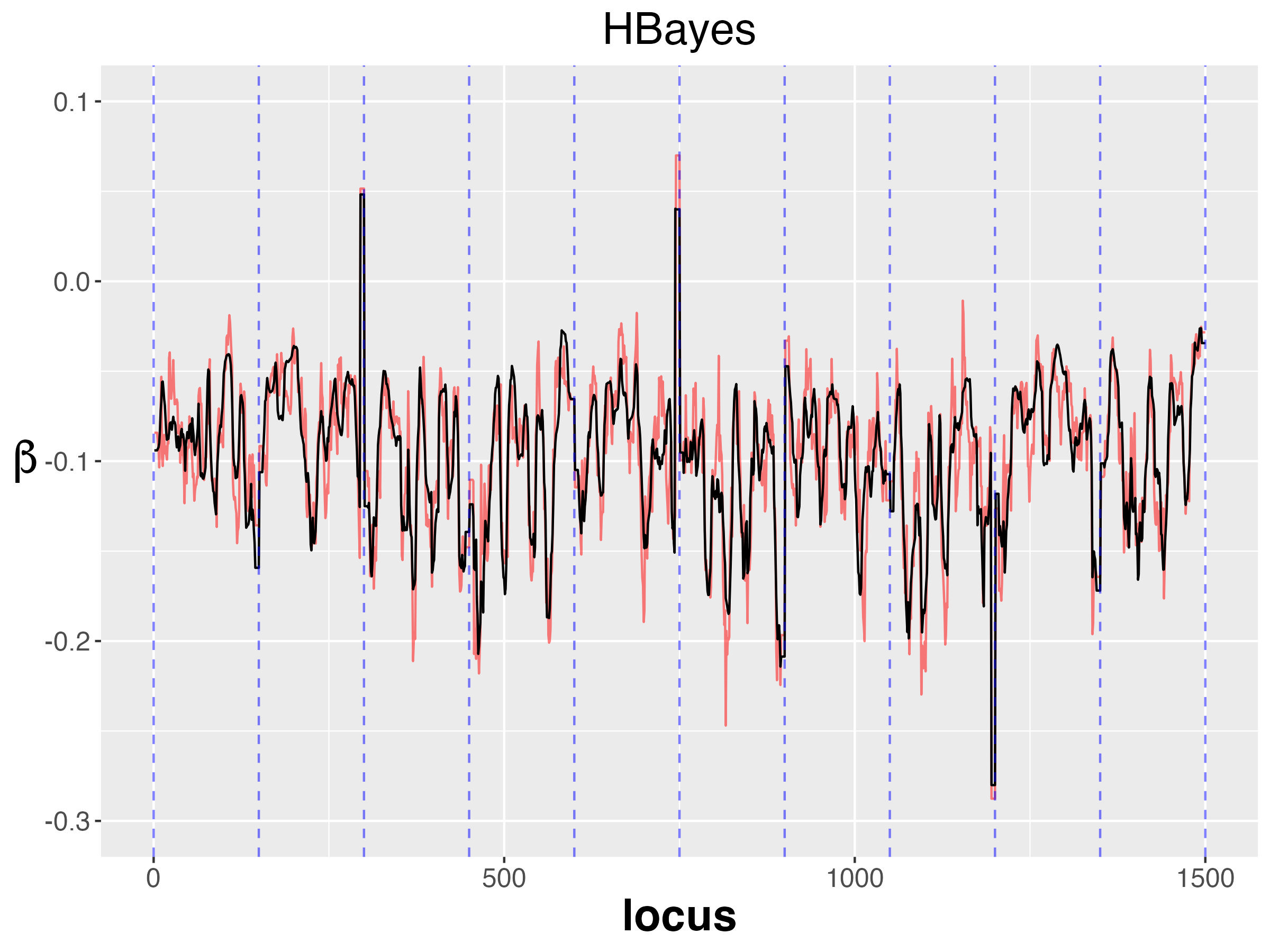}
\caption{Results of the analysis of one simulated data set. Going left to right, up to down we have: first plot is ridge, second is Mr.Ash of \cite{kim2022flexible}, third is the \cite{Wallin_22} method based on the mixed effect model, and fourth is our non-parametric Bayes approach. Red lines mark the true genetic effects and the black lines their estimates.
}
\label{fig:dense_simulation_2}
\end{figure}

Figure \ref{fig:sim_rmse_text} presents violin plots of RNE for our hierarchical Bayes method and some of the competitors mentioned above. 
% methods targeting dense signals, and \textit{Lasso} and \textit{EMVS}. 
The hierarchical Bayes method produces the best results in terms of estimation error, slightly outperforming the specialized method of \cite{Wallin_22}. 
Both of these methods yield the smallest MRNE while exhibiting appreciable variability when estimating regression coefficients at each of the very dense locations. 
This ``variability" effect disappears after smoothing the estimate over a 10 cM window. 
Interestingly, in terms of estimation error, ridge regression and \textit{Mr.Ash} are outperformed by \textit{EMVS} and cross-validation Lasso. 
As illustrated in Figure 
% 4 
\ref{fig:dense_simulation_2}, this is a result of oversmoothing of the signal in \textit{ridge} and \textit{Mr.~Ash}. 
However, in terms of prediction accuracy, ridge regression does well, while cross-validation Lasso and \textit{Mr.~Ash} are substantially worse, compared to other methods included in Figure \ref{fig:sim_rmse_text}.

\subsection{Real Data Analysis}
\label{subsec:real_data}
We use our hierarchical Bayes approach to analyze the popular  \cite{Z_00} {\it Drosophila} data. 
The purpose of the analysis is to identify genes influencing the shape of the posterior lobe of the male genital arch in {\it Drosophila}.
The size and shape variation of the males' posterior lobes (which are highly correlated) are quantified by averaging over both sides of the  morphometric descriptor (PC1) based on elliptical Fourier and principal components analyses.
These data were extensively analyzed in \cite{Z_00},
\cite{BF_08} and \cite{Wallin_22}, using different approaches based on the different multiple regression models. 
\cite{Z_00} and \cite{BF_08} use standard fixed effects multiple regression. 
\cite{Z_00} report 17 Quantitative Trait Loci (QTL), approximately uniformly distributed over the two chromosomes, with two of the strongest QTL located close to the centers of these chromosomes. 

\begin{figure}%[!ht]
 	\centering
 	\includegraphics[width=0.3\textwidth, height=3.5cm]{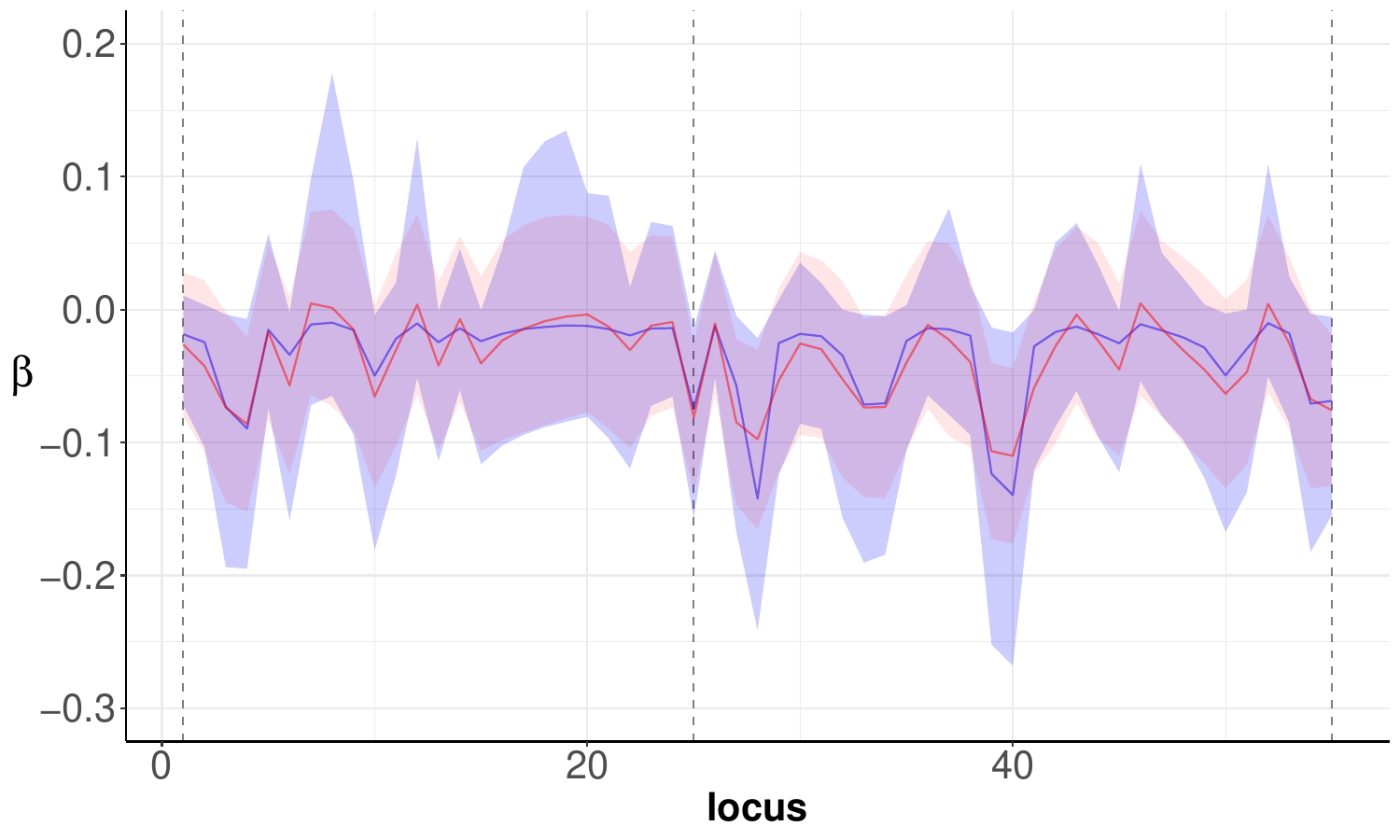}
 	\includegraphics[width=0.3\textwidth, height=3.5cm]{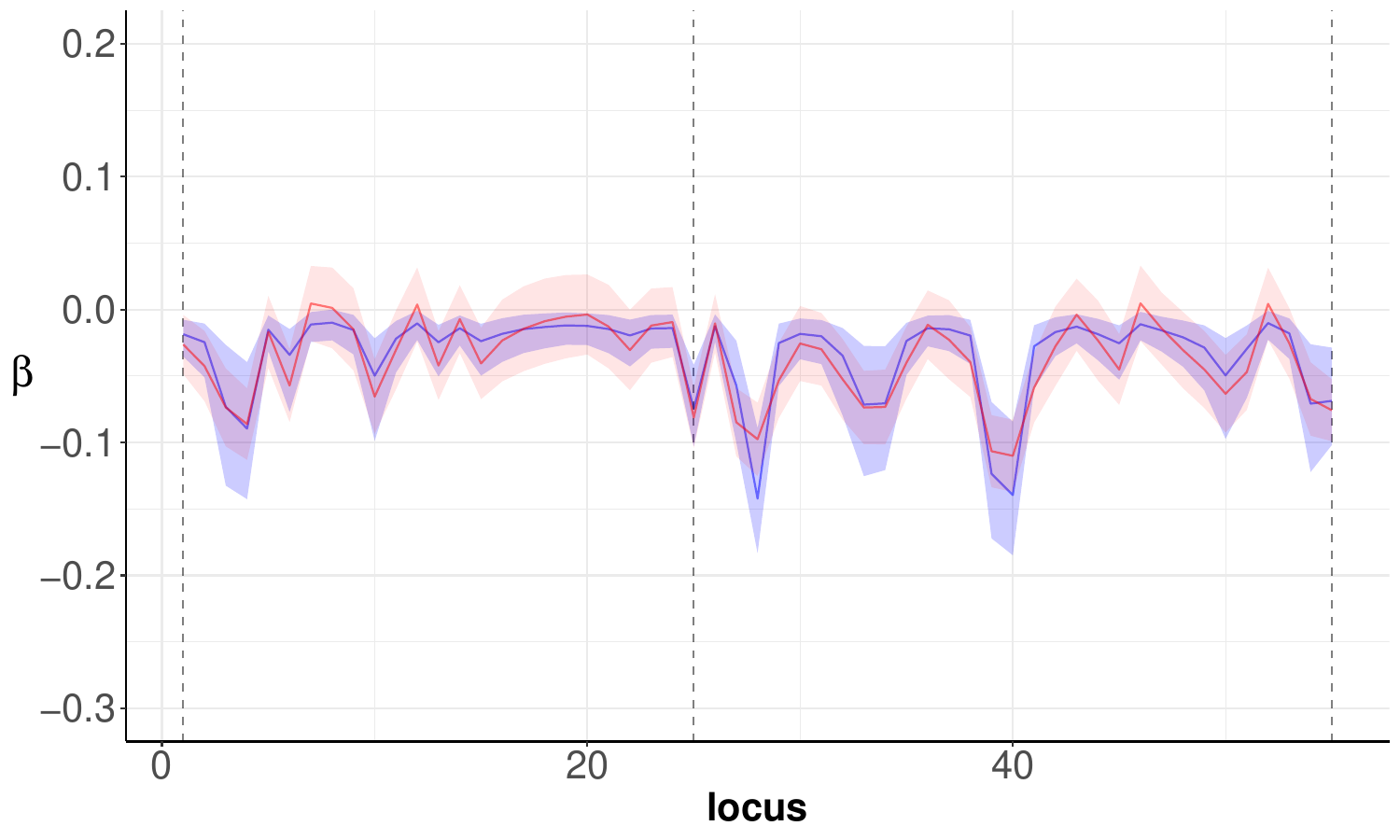}
         \includegraphics[width=0.3\textwidth, height=3.5cm]{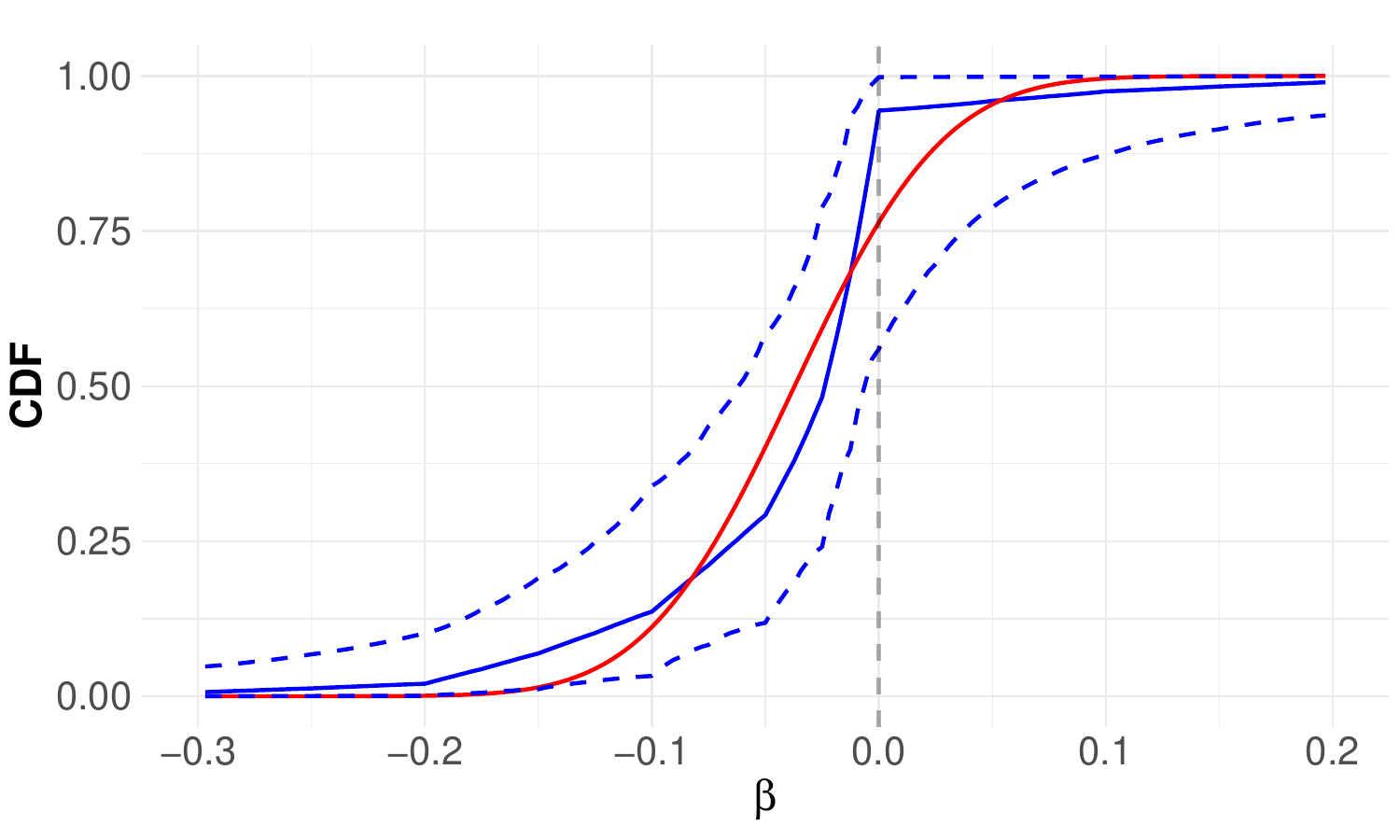}
         \vspace{-0.2cm}
         \caption{Two left panels are the posterior estimates of $\vec{\beta}$ ordered on the chromosomes. Blue and red areas represent pointwise credible intervals for our nonparametric Bayes method and the random effect model of \cite{Wallin_22}. The lines are corresponding posterior means. The plot to the left presents intervals at  $5\%-95\%$ coverage level, and
 	in the center, for level $25\%-75\%$. Right panel shows estimate of the CDF of the $\beta_j$'s for the random effects model (red) and the posterior median from the nonparametric Bayes method (blue). Dashed lines are pointwise $95\%$ credible intervals.} 
 \label{fig:real_betas_text}
 \end{figure}

The dataset  includes genotypes of 39 markers on 2 autosomes for $n=491$ individuals. 
Following \cite{BF_08}and \cite{Wallin_22}, we used  $m=161$ pseudo-marker explanatory variables spaced every 2cM.  
The values of these pseudo-markers are calculated  as the conditional expectations of the corresponding genotypes, given  the genotypes of observed flanking markers, as in the regression interval mapping of \cite{HK_92}. Such pseudo-marker explanatory variables are more strongly correlated than the markers spaced every 2cM. 
Thus, to curb the variance of locus specific estimates of regression coefficients, we performed our analysis using only every third of the pseudo-markers, i.e.~using pseudo-markers spaced every 6cM. 
Figure \ref{fig:real_betas_text} shows 95\% and 50\% Bayesian credible intervals for the loci-specific genetic effects and the estimates of the cumulative distribution of $\bfbeta$ for our nonparametric approach and the random effects model in  \cite{Wallin_22}.  

 Our analysis indicates a systematic negative polygenic effect, i.e.~many relatively weak QTL effects of the negative sign on both chromosomes. 
 The nonparametric Bayes estimates are more ``peaky'' than the estimates from a random effects model, especially  in the direction of negative values. 
 This difference is most visible in the plot of the 50\% credible intervals, which for the proposed method are entirely contained on the side of negative values. 
 Also, the median posterior distribution from the nonparametric Bayes approach is strongly asymmetric: it is almost truncated at zero, and has a relatively heavy negative tail. 
Heavy tails of the true polygenic distribution are also reflected in the 95\% credible intervals, which are wider for the nonparametric Bayes approach than for the mixed model of \cite{Wallin_22}. 
The Normal random effects model seems to over-smooth, and also places more weight on the positive values. 
Despite these differences, the summary results from both methods are quite similar. 
Both methods estimate heritability (\% of trait variability explained by genetic causes, $h^2= (\hat{\mathrm{V}}\left[Y\right]-\mathbb{E}\left[\sigma^2| {Y}\right]) \big/ \hat{\mathrm{V}}\left[Y\right]$) of  $72\%$, and the 95\% credible intervals for the posterior CDF of the polygenic effects cover the estimated Normal prior for the random effects model. 
Our approach gives more refined estimates  for locations with strong polygenic effects, and is able to capture asymmetry and heavy tails in the distribution.

\section{Discussion}\label{sec:discussion}
In this work we have proposed a new regularized estimator for high-dimensional GLMs. 
The method is inspired by an oracle shrinkage rule, defined as Bayes against the uniform prior on all  
permutations of the {\em true} vector $\bfbeta$. 
% Importantly, our oracle rule is a well defined object also in the frequentist setting, i.e., when $\vec{\beta}$ is fixed. 
The theory in Section \ref{sec:oracle} presents two different notions of optimality for the aforementioned oracle, a frequentist notion and a more general Bayesian notion, providing a formal justification for pursuing that oracle and motivating our nonparametric Bayes method (the relationship between the oracle and the data-driven hierarchical Bayes estimator is explained in Section 3). 
% In Section 3 we provided theoretical analysis on the optimality of the oracle shrinkage rule
% and explained the relation between the oracle Bayes and hierarchical Bayes estimators.
% We studied the performance of the estimators in simulations. 
The simulation results are viewed as proof-of-concept that the empirical distribution of the coefficients $\beta_j$ 
and the oracle Bayes and hierarchical Bayes estimators could be evaluated (approximated) effectively by Gibbs sampling algorithms. 
The simulations in Section \ref{sec:simulations}  consider fixed coefficient vectors and
asymptomatically independent $\vec{X}$ matrix, which, informally speaking, yield a model that is  approximately permutation invariant. 
% By extension of the arguments in the proof of Proposition \ref{prop:oracle-bayes-optimality-freq}, 
In this type of models the oracle Bayes rule actually minimizes {\em frequentist} risk (the result in Proposition \ref{prop:oracle-bayes-optimality-freq} under the Normal linear model being a special case). 
% the formal result for the Normal linear model is given 
% in Proposition 3.1. 
The simulation in Section \ref{subsec:real-data-simulation} which includes a non-exchangeable $\vec{X}$ matrix 
and a coefficient vector in which all but $3$ of the $1500$ components are i.i.d.~random samples, is in essence 
a Bayesian simulation that generates repeated joint realizations of the parameter and the data,  
and thus the average RNE entries reported in Table \ref{Table:RMSE_ICML} are practically evaluating {\em average} risks, and illustrate the result in Corollary \ref{cor1}. 
It is important to note that while our theoretical results apply to the {\em expectation} of the loss of the oracle Bayes estimator, in our simulations the oracle Bayes estimators actually yielded the smallest {\em losses},
and these were in turn well approximated by the proposed hierarchical Bayes estimates, 
for each simulation run.

\medskip
We conclude with some further comments on the relevance of the results in Section \ref{sec:oracle} to an analysis of the proposed method itself, which is 
an important direction for future work. 
% In Section \ref{sec:oracle} we presented the oracle $\widehat{\bfbeta}_{ol}$, which is Bayes w.r.t.~\eqref{eq:prior-oracle}, as pursuing a target defined in terms of the original frequentist model \eqref{eq:glm}. 
To formally analyze the proposed Polya tree-based method, one could start, as in \citet{antoniak1974mixtures}, with a genuinely Bayesian setup where $\beta_{j} \sim \Pi_{0}$, i.i.d., for some fixed and unknown distribution $\Pi_{0}$. 
In this case the oracle Bayes rule would be naturally defined as the minimizer of the Bayes risk under $\Pi_{0}$. 
If $P(\cdot |\boldsymbol{Y})$ denotes the posterior of $\Pi$ in our generative  model, which postulates {\em a priori} that $\Pi\sim P$ for a  Polya tree 
distribution $P$, then a primary goal would be to show that $P(\cdot \mid \boldsymbol{Y})$ converges to $\Pi_{0}$ under suitable conditions. 
If such a result can be obtained, it would generalize existing consistency results on density estimation with Polya tree priors \citep[e.g., ][]
{castillo2017polya}. 
If $P(\cdot \mid \boldsymbol{Y})$ converges to $\Pi_{0}$, which in terms of the representation in \eqref{eq:gen:pst} means that the posterior of $\boldsymbol{\phi}$ converges to some $\boldsymbol{\phi}_0$ (corresponding to  $\Pi_0$), then the Bayes rule for $\boldsymbol{\beta}$ under our generative  model should in turn be consistent for the Bayes rule under the true prior $\Pi_{0}$. 
Existing consistency results in the literature, for example those in \citet{castillo2017polya}, correspond to the situation where the $\beta_{j}$'s are observed. 
% , and are already nontrivial to prove; establishing consistency 
In 
our case, where only the $Y_i$'s are observed, and the likelihood of each $Y_i$ depends on the common vector parameter  $(\beta_{1}, \ldots, \beta_{p})$, is considerably more challenging and deserves 
separate consideration. 

\section*{Acknowledgments}
A.W. was supported by the Israeli Science Foundation (ISF) under grant no.~2679/24. 
M.B.~and J.W.~were supported by the Swedish Research Council under grant no.~2020-05081. 

% \bibliographystyle{apalike} % Or "plainnat", but JCGS prefers author-year
% \bibliography{yourbibfile}  % Your bibliography .bib file

\bibliographystyle{apalike}
\bibliography{refs}

\appendix

\section{Proofs}
\begin{proof}[Proof of Proposition \ref{prop:oracle-bayes-optimality-freq}]
Let $\widehat{\bfbeta}(\bZ)$ be any PI rule under the (PI) model \eqref{eq:model-suff}. 
Then we can proceed as in \cite{weinstein2021permutation} and calculate the risk of $\widehat{\bfbeta}$ at $\bfbeta^*$ as 
\begin{equation}\label{eq:risk-pi}
\begin{aligned}
R(\bfbeta^*, \widehat{\bfbeta}) &= \EE_{\bfbeta^*} L(\bfbeta^*, \widehat{\bfbeta}(\bZ))\\ 
&= \EE_{\bfbeta^*} L(\tau(\bfbeta^*), \tau(\widehat{\bfbeta}(\bZ)))\\
&=\EE_{\bfbeta^*} L(\tau(\bfbeta^*), \widehat{\bfbeta}(\tau(\bZ)))\\
&=\EE_{\tau(\bfbeta^*)} L(\tau(\bfbeta^*), \widehat{\bfbeta}(\bZ))\\
&=R(\tau(\bfbeta^*), \widehat{\bfbeta}),
\end{aligned}
\end{equation}
and we remind that the subscript on the expectation operator is the value of the parameter indexing the distribution of $\bZ$ (not of $\tau(\bZ)$). 
Above, the second equality is because the {\em loss} is PI, the third equality is because the {\em rule} $\widehat{\bfbeta}$ is PI, and, crucially, the fourth inequality is because the {\em model} for $\bZ$ is PI under \eqref{eq:gram}. 
From \eqref{eq:risk-pi} it follows that 
$$
R(\bfbeta, \widehat{\bfbeta}) = \frac{1}{p!}\sum_{\tau} R(\tau(\bfbeta^*), \widehat{\bfbeta}),
$$
the sum taken over all $p!$ permutations $\tau$. 
But this is precisely the Bayes risk of $\widehat{\bfbeta}$ under the prior $\widetilde{\Pi}^*_p$. 
The proof is complete because the oracle Bayes rule is {\em defined} to be the Bayes rule under the prior $\widetilde{\Pi}^*_p$. 
\end{proof}

\begin{proof}[Proof of Proposition \ref{prop:oracle-bayes-optimality}]

Let $\widetilde{\Pi}$ be any exchangeable prior on $\bfbeta$. 
Per the technical modification in the statement of the proposition, the oracle Bayes rule $\widehat{\bfbeta}_{ol}$ is now also a function the true (random) parameter vector $\bfbeta$ (through $\{\bfbeta\}$). 
Thus, fist note that, quite trivially, 
\begin{equation}\label{eq:bayes-general}
\min_{{\widehat{\bfbeta}}} \ \mathbb{E}_{\widetilde{\Pi}}[\ L(\ \bfbeta, \ \widehat{\bfbeta}(\bY,\{\bfbeta\})\ ) \ ] \leq \min_{\widehat{\bfbeta}} \ \mathbb{E}_{\widetilde{\Pi}}[\ L(\ \bfbeta, \ \widehat{\bfbeta}(\bY)\ )\ ], 
\end{equation}
where on the left hand side the minimum is over all functions $\widehat{\bfbeta}$ of $(\bY,\{\bfbeta\})$, and on the right hand side the minimum is over all functions $\widehat{\bfbeta}$ of $\bY$ only; and where in both sides of the inequality the expectation is with respect to the joint distribution of $(\bY,\bfbeta)$ under the prior $\widetilde{\Pi}$. 
% Indeed, since any function of $\bY$ only is also a function of $(\bY,\{\bfbeta\})$, the minimum on the left hand side of \eqref{eq:bayes-general} is taken over a larger set of rules. 
Therefore, it is enough to show that the oracle Bayes rule $\widehat{\bfbeta}_{ol}$ minimizes the left hand side of \eqref{eq:bayes-general}, i.e., that 
\begin{equation}\label{eq:pf-bayes}
\argmin_{\bfb\in \RR^p}\EE_{\widetilde{\Pi}} [L(\bfbeta, \bfb) \lvert \bY, \{\bfbeta\}] = 
\argmin_{\bfb\in \RR^p}\EE_{\widetilde{\Pi}^*} [L(\bfbeta, \bfb) \lvert \bY ],
\end{equation}
Now, the posterior of $\bfbeta$ on the left hand side of \eqref{eq:pf-bayes} 
is supported on the set of all possible orderings of the components of $\{\bfbeta\}$, i.e., on all permutations of $\bfbeta$. 
Calculating the posterior for $\bfbeta$ of $\bfbeta$, we have 
\begin{equation}\label{eq:pf-bayes-2}
\widetilde{\Pi}(\bfbeta|\bY,\{\bfbeta\})\propto \Pi(\bfbeta|\{\bfbeta\})f(\bY|\bfbeta, \{\bfbeta\}) = \widetilde{\Pi}(\bfbeta|\{\bfbeta\})f(\bY|\bfbeta),
\end{equation}
and, since $\widetilde{\Pi}$ is exchangeable, 
\begin{equation}\label{eq:pf-bayes-3}
\widetilde{\Pi}(\bfbeta|\{\bfbeta\}) = \widetilde{\Pi}^*, 
\end{equation}
the uniform distribution on all permutations of $\bfbeta$. 
From \eqref{eq:pf-bayes-2} and \eqref{eq:pf-bayes-3}, we conclude that the posterior of $\bfbeta$ given $\bY$ and $\{\bfbeta\}$ is exactly the posterior with respect to which the minimum in \eqref{eq:oracle-bayes} is taken. 
This completes the proof. 
\end{proof}

% \section{Gibbs sampling of $\beta$}\label{sec:sample_beta}

\section{Gibbs sampling of \texorpdfstring{$\boldsymbol{\beta}$}{beta}}\label{sec:sample_beta}

To sample the vector $\bfbeta = (\beta_1,...,\beta_p)$ given $(\psi, \boldsymbol{\phi}, \boldsymbol{Y})$, we utilize a MH-within-Gibbs algorithm inspired by the coordinate descent algorithm of \cite{friedman2007pathwise}, that was shown to work very well for Lasso. 
Let $\boldsymbol{r} = (r_1,...,r_p )$ denote the  index 
of the  FPT subintervals to which  each component of $\bfbeta$ belongs, i.e.~$r_j= k$ if $\beta_j \in {\cal I}_{L,k}$. 
We now detail how to generate a proposal for the MH algorithm for $\beta_j$:
\begin{enumerate}
	\item Draw $r^* \sim r_j + Unif\{-K,\ldots,K\}$, where $K$ is a parameter which we estimate using an adaptive MCMC (AMCMC) scheme similar to \cite{roberts2009}. 
	Specifically, we take an increasing  batch size of MCMC samples; if the acceptance rate for the MH algorithm is above $0.3$ we increase $K$, if it is below $0.3$ we decrease it. 
	The batch size is chosen such that updating is more seldom, in order to ensure convergence of the AMCMC algorithm. 
	\item Obtain a Taylor approximation of the log-likelihood $l(\beta_j)  = \log  f \left(\boldsymbol{Y}|\vec{X},\vec{\beta},\psi \right)$, 
	$$
	l(\beta^*_j) \approx l(\beta_j)   + l'(\beta_j)  \left(\beta^*_j- \beta_j\right) + \frac{l''(\beta_j)}{2 }\left(\beta^*_j- \beta_j\right)^2,
	$$
	where 
	$$
	l'(\beta_j) = \frac{\partial}{\partial \beta_j} \log  f\left(\boldsymbol{y}|\vec{\beta},\psi   \right)\ \ \  \ \ \text{and}\ \ \  \ \ l''(\beta_j) = \frac{\partial^2}{\partial \beta^2_j} \log f\left(\boldsymbol{Y}|\bfbeta,\psi  \right). 
	$$
	
	\item Use the Taylor approximation to obtain a Normal approximation of the posterior as a proposal distribution, then generate a sample from the proposal given that proposal is contained in ${\cal I}_{L,r^*}$. 
	That is, $\beta^*_j \sim \mathcal{N}_{{\cal I}_{L,r^*}}\left(\mu,\sigma^2\right)= \mathcal{N}\left(\mu, \sigma^2 \right)| \{\beta^*_j \in {\cal I}_{L,r^*}\}$, where $ {\cal I}_{L,r^*} = (a_{r^*},a_{r^*+1}]$, $\mu=\beta_j  + \frac{l'(\beta_j) }{-l''(\beta_j)}$, and $\sigma^2=  \frac{1}{-l''(\beta_j)}$.%\swarrow\swarrow
\end{enumerate}
Thus, we first generate an interval ${\cal I}_{L,r^*}$ that includes $\beta^*_j$, then, given that the prior is constant on the interval, we use a quadratic approximation of the likelihood to sample $\beta^*$ given it is in $ {\cal I}_{L,r^*}$. 
Exactly as with coordinate descent algorithms, the main advantage of the algorithm is computational efficiency, for instance, one does not need to compute $\boldsymbol{X}\bfbeta$ in each iteration, but instead store $\hat{\boldsymbol{Y}} = \boldsymbol{X}\bfbeta$ and then compute 
% \revv{
$\hat{\boldsymbol{Y}}^*  = \hat{\boldsymbol{Y}} + ( \beta^*_j - \beta_j )\bX^{(j)}$, where $\bX^{(j)}$ is the $j$th column of $\bX$. 
% It seems that $\beta_j,\beta_j^*$ are scalars while $\bx_j$ is a vector (right?). Did you mean the vectors $(\beta_1,...,\beta_p)$ and $(\beta_1^*,...,\beta_p^*)$?}
% }
Hence, instead of computing a matrix-vector multiplication we only carry out a vector-scalar multiplication, and vector-addition.  
For the general likelihood \eqref{eq:glm}, note that the log-likelihood, the gradient and the Hessian can be computed using only  $\boldsymbol{X}\boldsymbol{\beta}$, and $\bX^{(j)}$ .

\subsection{Sampling with oracle prior}\label{sec:sample_oracle}
Here we describe how to sample using MCMC when when the prior is the oracle prior $\pi_0$ defined in \eqref{eq:prior-oracle}. Since the  vector of unique coefficients are fixed we simple permute the locations randomly. 
To generate proposal $\bfbeta^*$ given a previous sample $\bfbeta$ we do as follows: first set $\bfbeta^*= \bfbeta$, second  generate two indices uniformly $i_1,i_2 \sim Unif\{1,2,\ldots,p\}$ and set 
$\left(\beta^*_{i_1} , \beta^*_{i_2} \right) = \left( \beta_{i_2},  \beta_{i_1} \right)$. 
Clearly this proposal is symmetric, and since oracle prior is constant over the permutation, the Metropolis Hastings ratio is just $\frac{f(\bY;\bfbeta^*,\psi)}{f(\bY;\bfbeta,\psi)}$.

\end{document}